\documentclass[a4paper,12pt]{article}
\usepackage{latexsym,amssymb}
\usepackage{graphicx}
\usepackage{cite}
\usepackage[arrow,matrix,curve]{xy}
\usepackage{amsmath}

\global\arraycolsep=2pt 
\input{epsf} 

\def\s[#1,#2]{[#1\stackrel{{\displaystyle\star}}{,}#2]}

 1

\usepackage{amsthm}

\newtheorem{Definition}{Definition}

\newcommand{\eq}{\begin{equation}}
\newcommand{\eqa}{\begin{eqnarray}}
\newcommand{\en}{\end{equation}}
\newcommand{\ena}{\end{eqnarray}}
\newcommand{\enn}{\nonumber \end{equation}}

\def\sk{\vskip .4cm}
\def\noi{\noindent}

\def\de{\delta}
\def\epsi{{\varepsilon}}
\def\st {\star}
\def\f{{\rm{f}\,}}
\def\of{{\overline{{\rm{f}}\,}}}

\def\D/h{\widehat{\fmslash D}}

\def\om{\omega}
\def\Om{\Omega}
\def\al{\alpha}
\def\la{\lambda}
\def\be{\beta}
\def\ga{\gamma}
\def\Ga{\Gamma}

\def\de{\delta}

\def\5bar{{\overline 5}}

\def\RR{{\mathcal R}}

\def\R{{R}}
\def\oR{{\overline{\R}}}

\def\UU{{U}\Xi}

\def\FF{\mathcal F}

\def\varepsi{\varepsilon}

\def\s'O{\stackrel{_{{\displaystyle\st \footnotesize '}}}{_{^{^{\displaystyle\otimes}}}}}

\def\vphi{\varphi}

\def\AAs{{A_\st }}
\def\UUs{{\UU_\st }}
\def\Xis{{\Xi_\st }}
\def\Oms{\Omega_\st}
\def\La{\Lambda }

\def\ll{{\mathcal L}}
\def\D{\Delta}
\def\1s{{1_\st }}
\def\3s{{3_\st }}
\def\2s{{2_\st }}
\def\ef1{{1_\FF}}
\def\ef2{{3_\FF}}
\def\ef3{{2_\FF}}

\def\TT{{\mathcal T}}
\def\lls{\ll^\st }

\def\hbar{\lambda}
\def\Diff{\mbox{\it Diff}\,}

\def\le{\langle}
\def\re{\rangle}
\def\nn{\nonumber}

\def\bbC{\mathbb{C}}
\def\AA{\mathcal A}
\def\dd{{\triangledown}}
\def\dds{\triangledown^\st}

\def\rr{\mathsf{R}}
\def\ric{\mathsf{Ric}}
\def\tr{\mathsf{T}}

\def\g{\mathsf{g}}

\def\ots{\otimes_\st}
\def\res{\re_\st}
\def\CC{\mathcal{C}^\st}
\def\ste{*}

\numberwithin{equation}{section}

\begin{document}

\begin{titlepage}
\rightline{DISTA-UPO/05}
\rightline{LMU-ASC 66/05}
\rightline{MPP-2005-119}
\sk\sk\sk\sk
\begin{center}
{\bf\LARGE{Noncommutative Geometry and Gravity}}
\vskip 3em

{{\bf Paolo Aschieri${}^{1}$, Marija Dimitrijevi\' c${}^{2,3,4}$, 
Frank Meyer${}^{2,3, \dagger}$ and Julius Wess${}^{2,3}$ }}

\vskip 1.5em

${}^{1}$Dipartimento di Scienze e Tecnologie Avanzate\\
Universit\' a del Piemonte Orientale, and INFN\\
Via Bellini 25/G 15100 Alessandria, Italy\\[1em]

${}^{2}$Arnold Sommerfeld Center for Theoretical Physics\\
Universit\"at M\"unchen, Fakult\"at f\"ur Physik\\
Theresienstr.\ 37, 80333 M\"unchen, Germany\\[1em]

${}^{3}$Max-Planck-Institut f\"ur Physik\\
F\"ohringer Ring 6, 80805 M\"unchen, Germany\\[1em]

${}^{4}$University of Belgrade, Faculty of Physics\\
Studentski trg 12, 11000 Beograd, Serbia and Montenegro\\[1em]

\end{center}

\sk
\sk
\centerline{\bf Abstract}
\sk
\normalsize{We study a deformation of infinitesimal diffeomorphisms of a
smooth manifold. The deformation is based on a general twist.
This leads to a differential geometry on a noncommutative algebra of 
functions whose product is a star-product. The class of noncommutative
spaces studied is very rich. 
Non-anticommutative superspaces are also briefly considered.

The differential geometry developed is covariant under deformed
diffeomorphisms and it is coordinate independent. The main target
of this work is the construction of Einstein's equations for gravity 
on noncommutative manifolds.}

\sk\sk
\footnotesize{\noi PACS: 02.40.Gh, 02.20.Uw, 04.20.-q,  11.10.Nx, 04.60.-m.~~~ 
2000 MSC: 83C65, 53D55, 81R60, 58B32\\[.17cm]
\noi E-mail: aschieri,dmarija,meyerf,wess@theorie.physik.uni-muenchen.de}\\

\end{titlepage}\vskip.2cm

\newpage

\tableofcontents
\section{Introduction}




The study of the structure of spacetime at Planck scale,
where quantum gravity effects are non-negligible, 
is one of the main open challenges in fundamental 
physics. Since the dynamical variable in Einstein general relativity  
is spacetime itself (with its metric structure), and since in quantum
mechanics and in quantum field theory the classical dynamical variables 
become noncommutative, one  is strongly lead to conclude that 
noncommutative spacetime is a feature of Planck scale physics. 
   This expectation is further supported by Gedanken experiments 
that aim at probing spacetime structure at very small distances. 
They show that due to gravitational backreaction one cannot test 
spacetime at Planck scale\footnote{For example, 
in relativistic quantum mechanics the position of a particle can
be detected with a precision at most of the order of its 
Compton wave length $\lambda_C=\hbar/mc$.
Probing spacetime at infinitesimal distances implies an extremely 
heavy 
particle that in turn curves spacetime itself. 
When $\lambda_C$ is of the order of the Planck length, the spacetime
curvature 
radius due to the particle has the same order of magnitude and the 
attempt to measure spacetime structure beyond Planck scale fails.}. 
Its description as a (smooth) manifold becomes therefore  
a  mathematical assumption no more justified by physics. It is then natural to 
relax this assumption and conceive a more general noncommutative spacetime, 
where uncertainty relations and discretization naturally arise. In this way one
can argue  for the impossibility of an operational definition of 
continuous Planck 
lenght spacetime (i.e., a definition given by describing the operations to be 
performed for at least measuring spacetime by a Gedanken experiment). A 
dynamical 
feature of spacetime could be incorporated at a deeper kinematical 
level. As an example compare 
Galilean relativity to special relativity. Contraction of distances and 
time dilatation can be explained in Galilean relativity: 
they are a consequence of the interaction between ether and the 
body in motion. In special relativity they have become a kinematical feature.

This line of thought has been pursued in previous works, starting with
\cite{Pauli:1985},\cite{Snyder:1946qz},
and more recently in  \cite{Madore:1992} -\cite{Calmet:2005qm}
.

Notice that uncertainty relations in position measurements are also 
in agreement with string theory models \cite{Veneziano}.  
Moreover, non-perturbative attempts to 
describe string theories have shown that a noncommutative structure of 
spacetime emerges
\cite{Banks}. 

\sk
A first question to be asked in the context we have outlined
is whether one can consistently deform Riemannian geometry into 
a noncommutative Riemannian geometry.
We address this question by considering deformations of the 
algebra of functions on a manifold obtained via a quite 
wide class of $\st$-products. 
In this framework we successfully construct
a noncommutative version of differential and of Riemannian geometry,
and we obtain the noncommutative version of Einstein equations.
\sk
Even without physical motivations, the mathematical structure of
deformed spaces is a challenging and fruitful research arena.
It is very surprising how well $\st$-noncommutative structures can be 
incorporated in the framework of differential geometry.

\sk
The $\st$-products we consider are associated with a deformation by a 
twist $\FF$ of the Lie algebra of infinitesimal diffeomorphisms on
a smooth manifold $M$. Since $\FF$ is an arbitrary twist, we can consider it as the dynamical variable that determines the possible
noncommutative structures of spacetime.
\sk
In Section 2 we construct the universal enveloping algebra $\UU$ of the 
Lie algebra of vectorfields, and we give a pedagogical description 
of its Hopf algebra structure. The twists we consider are elements 
$\FF\in \UU\otimes\UU$.
The notion of twist of a Lie algebra is well 
known \cite{Drinfeld1, Drinfeld3}. 
Multiparametric twists appear in \cite{Reshetikhin}.
Other examples of twists (Jordanian deformations)
are in \cite{GER}, \cite{OGIEV}
and \cite{Kulish1}.
In the context of deformed Poincar\'e group and Minkowski space geometry 
twists have been studied in \cite{ACPoincare}, \cite{Kulish-Mudrov} 
(multiparametric deformations),
and in \cite{Wess}, \cite{Chaichian}, \cite{Koch} 
\cite{Gonera:2005hg}, \cite{Oeckl:2000eg}
(Moyal-Weyl deformations), 
see also \cite{Lukierski}.

In the context of Connes noncommutative geometry, 
the noncommutative  torus, 
the noncommutative spheres \cite{Connes-Landi} and 
further noncommtative manifolds (so-called isospectral 
deformations) considered in \cite{Connes-Landi}, 
and in \cite{Dubois-Violette}, 
\cite{Gayral:2005ad}, 
are  noncommutative manifolds whose deformed algebra of functions 
is along the lines of Rieffel's twists \cite{Rieffel}; see
\cite{Sitarz} and, for the four-sphere in \cite{Connes-Landi},
see \cite{Varilly}, \cite{Aschieri-Bonechi}. 

Our contribution in this section is to consider the notion 
of twist in  the context of an infinite dimensional Lie algebra, 
that of vectorfields on $M$.
Several examples of twists and of their corresponding 
$\st$-noncommutative 
algebra of funtcions are then presented. We also extend
this notion to the case where $M$ is superspace, 
and describe in a sound mathematical setting a very general class 
of twists on superspace. 

We conclude Section 2 by recalling the construction of the Hopf algebra
$\UU^\FF$ \cite{Drinfeld3}. 
This Hopf algebra is closely related to the 
Hopf algebra of deformed infinitesimal diffeomorhisms.

\sk
We begin Section 3 by recalling some known facts about 
Hopf algebra representations and then construct the algebra $\UUs$
(with product $\st$) as a module algebra on which $\UU^\FF$ acts. 
The space of vectorfields has a deformed Lie bracket that is realized as 
a  deformed commutator in $\UUs$. We have constructed the deformed 
Lie algebra of infinitesimal diffeomorphisms
(infinitesimal $\st$-diffeomorphisms).
We then construct a natural Hopf algebra 
structure on $\UUs$ which proves that vectorfields form a
deformed Lie algebra in the sense of \cite{Woronowicz}, 
see also \cite{AC}, \cite{SWZ}, and \cite{AschieriTesi} p.~41. 
It can also be proven that $\UU_\st$ and $\UU^\FF$ are isomorphic Hopf algebras
\cite{NG}. In \cite{G1}, \cite{Dimitrijevic}
and \cite{Meyer} (where $\theta^{\mu\nu}$-constant noncommutativity is
considered) the Hopf algebra $\UU^\FF$
rather than $\UU_\st$ is used.

\sk
In Section 4 we study the $\st$-action of the Hopf algebra of infinitesimal 
$\st$-diffeomorphisms on the algebra of noncommutative functions 
$\AAs\equiv Fun_\st(M)$ and on $\UUs$. In the same way that $A\equiv 
Fun(M)$ and 
$\UU$ were deformed in Section 3, we here deform the algebra of tensorfields 
$\TT$ into $\TT_\st$ and then study the action of $\st$-diffeomorphisms
on $\TT_\st$. As a further example we similarly proceed with the algebra
of exterior forms. 

We then study the pairing between vectorfields and 1-forms, and its  
$\AAs$-linearity properties. Moving and dual comoving frames (vielbein)
are introduced. As in the commutative case, (left) $\AAs$-linear maps 
$\Xis\rightarrow \AAs$ are the same as  1-forms. More in general tensorfields
can be equivalently described as (left) $\AAs$-linear maps.  
\sk
In Section 5 we define the $\st$-covariant derivative in a global coordinate 
independent way. Locally the covariant derivative is completely determined by 
its coefficients $\Ga_{\mu\nu}^\sigma$. Using the deformed Leibniz rule for 
vectorfields we extend the covariant derivative to all type of tensorfields.

In Section 6 torsion, curvature and the Ricci tensors are defined as 
(left) $\AAs$-linear maps on vectorfields. The $\AAs$-linearity property is a 
strong requirement that resolves the ambiguities in the possible definitions of
these noncommutative tensorfields.

In Section 7 we define the metric as an arbitrary $\st$-symmetric element 
in the $\st$-tensorproduct of 1-forms $\Oms\ots\Oms$. Using the pairing 
between vectorfields and 1-forms the metric is equivalently described as an
$\AAs$-linear map on vectorfields, $(u,v)\mapsto \g(u,v)$. 
The scalar curvature is then defined and Einstein equations on 
$\st$-noncommutative space are obtained. Again the requirement of 
$\AAs$-linearity uniquely fixes the possible ambiguities arising
in the noncommutative formulation of Einstein  gravity theory.

In Section 8 we study reality conditions on noncommutative 
functions, vectorfields and tensorfields. If the 
twist $\FF$ satifies a mild natural extra condition 
then all the geometric  constructions achieved in the previous sections 
admit a real form.

\section{Deformation by twists}

\subsection{Hopf algebras from Lie algebras}\label{SEct2.1}

Let us first recall that the (infinite dimensional) linear space $\Xi$ of smooth vectorfields on a smooth manifold $M$ becomes a Lie algebra through the map
\begin{eqnarray}
[\quad ]: \quad\quad \Xi\times\Xi &\to& \Xi \nonumber\\
(u,v) &\mapsto& [u~v] .\label{2.1}
\end{eqnarray}
The element $[u~v]$ of $\Xi$ is defined by the usual Lie bracket
\begin{equation}
[u~v](h) = u(v(h)) - v(u(h)) .\label{2.2}
\end{equation}
We shall always denote vectorfields by the letters $u$, $v$,
$z$,\dots and functions on $M$ by $f$, $g$, $h$,\dots. 

The Lie algebra of vectorfields (i.e. the algebra of infinitesimal diffeomorphisms) can also be seen as an abstract Lie algebra without referring to the smooth manifold $M$ anymore. This abstract algebra can be extended to a Hopf algebra by first defining the universal enveloping algebra $\UU$ that is 
the tensor algebra (over $\mathbb{C}$) generated by the elements of $\Xi$ 
and the unit element $1$ modulo the left and right ideal generated by 
all elements 
$uv-vu-[u~v]$. The elements $uv$ and $vu$ are 
elements in the tensor algebra  and $[u~v]$ is an element of $\Xi$. We shall denote elements of the universal enveloping algebra $\UU$ by $\xi$, $\zeta$, $\eta$,\dots. 

The algebra $\UU$ has a natural Hopf algebra structure \cite{Sweedler,Majidbook}.
On the generators $u\in \Xi$ and the unit element $1$ we define
\eqa\label{cosclass}
&&\Delta (u)=u\otimes 1 + 1\otimes u\nonumber~,~~~~~\Delta(1)=1\otimes 1~,\\
&&\varepsi(u)=0~~,~~~~~~~~~~~~~~~~~~~~~\varepsi(1)=1~,\\
&&S(u)=-u\nonumber~~,~~~~~~~~~~~~~~~~~~S(1)=1~.
\ena
Here $\Delta $ is the coproduct (from which the Leibniz rule for vectorfields follows),
$S$ is the antipode (or coinverse) and $\epsi$ the counit.
The maps $\Delta$, $\varepsilon$ and $S$  satisfy the
following relations
\begin{eqnarray}
\Delta (u)\Delta (v) - \Delta (v)\Delta (u) &=& [u~v]\otimes 1 + 1\otimes [u~v] = \Delta([u~v])~, \nonumber\\
\varepsilon(u)\varepsilon(v) - \varepsilon(v)\varepsilon(u) &=& \varepsilon ([u~v])~ , \nonumber\\
S(v)S(u) - S(u)S(v) &=& vu-uv = S([u~v])~. \label{2.4}
\end{eqnarray}
This allows us to extend $\Delta$ and  $\varepsilon$ 
as algebra homomorphisms
and $S$ as antialgebra homomorphism to the full enveloping algebra,
$\Delta: \UU\rightarrow \UU\otimes \UU$, 
$\varepsi:\UU\rightarrow \mathbb{C}$ 
and $S:\UU\rightarrow \UU$, 
\begin{eqnarray}
\Delta (\xi\zeta) &:=& \Delta(\xi)\Delta(\zeta) ,\nonumber\\
\varepsilon (\xi\zeta) &:=& \varepsilon(\xi)\varepsilon(\zeta) ,\nonumber\\
S(\xi\zeta) &:=& S(\zeta)S(\xi) .\label{2.5}
\end{eqnarray}
There are three more propositions that have to be satisfied for a Hopf 
algebra (we denote by $\mu$ the product in the algebra)
\begin{eqnarray}
(\D \otimes id)\D(\xi) &=& (id \otimes \D)\D(\xi)~ ,\nonumber\\
(\varepsilon \otimes id)\D(\xi) &=& (id \otimes \varepsilon)\D(\xi) 
= \xi~ .\nonumber\\
\mu(S\otimes id)\D(\xi) &=& \mu(id \otimes S)\D(\xi) = \epsi(\xi)1 ~,\label{2.6}
\end{eqnarray}
It is enough  to prove (\ref{2.6}) on the generators $u,
1$ of $\UU$.
We prove the first of them for the coproduct defined in (\ref{cosclass}) using the Sweedler notation 
$\Delta(u) = u_1 \otimes u_{2}$ (where a sum over $u_1$ and $u_2$ is
understood), in this explicit case  $\Delta(u)=u_1 \otimes u_{2}=u\otimes 1+
1\otimes u\,$,
\begin{eqnarray}
(\D \otimes id)\D(u) &=& \D (u_1) \otimes u_2 \nonumber \\
& = &u_{1_1}\otimes u_{1_2}\otimes u_2 \nonumber\\
&=& (u\otimes 1 + 1\otimes u)\otimes 1 + 1\otimes 1\otimes u \label{2.7}
\end{eqnarray}
and
\begin{eqnarray}
(id \otimes \D)\D(u) &=& u_{1} \otimes \Delta (u_{2}) \nonumber \\
&=& u_1\otimes u_{2_1}\otimes u_{2_2} \nonumber\\
&=& u\otimes 1\otimes 1 + 1\otimes(u\otimes 1 + 1\otimes u) .\label{2.8}
\end{eqnarray}
Comparing (\ref{2.7}) and (\ref{2.8}) we see that the first condition of (\ref{2.6}) is satisfied.

After proving the remaining conditions of (\ref{2.6}) on the
generators of $\UU$
we have constructed the Hopf algebra 
$(\UU,\cdot, \Delta, S,\varepsi)$, where $\cdot$ denotes the
multiplication map in $\UU$; sometimes we denote it by
$\mu$ and frequently omit any of the symbols $\cdot$ and $\mu$. With abuse of notation we frequently write $\UU$ to denote the Hopf algebra $(\UU,\cdot, \Delta, S,\varepsi)$.
This Hopf algebra is cocommutative because $\Delta=\Delta^{op}$
where $\Delta^{op}=\sigma \circ \Delta$ with $\sigma $ the flip map 
$\sigma(\xi\otimes\zeta)=\zeta\otimes\xi$.  
\sk
We will extend the notion of
enveloping algebra to formal power series in $\hbar$, and we will 
correspondingly consider the Hopf algebra $(\UU[[\lambda]],\cdot,\Delta, S,\varepsi)$.
In the sequel for sake of brevity we will often denote 
$\UU[[\la]]$ by $\UU$.

\subsection{The twist}\label{THETWIST}

\begin{Definition}\label{twistdeff}
A twist $\FF$ is an element 
$\FF\in \UU[[\la]]\otimes \UU[[\la]]$ that is invertible and that satisfies
\eq\label{propF1}
\FF_{12}(\Delta\otimes id)\FF=\FF_{23}(id\otimes \Delta)\FF\,,
\en
\eq\label{propF2}
(\varepsi\otimes id)\FF=1=(id\otimes \varepsi)\FF~,
\en
where $\FF_{12}=\FF\otimes 1$ and $\FF_{23}=1\otimes \FF$.
\sk
\end{Definition}
In our context we in addition require\footnote{ Actually it is
possible to show that (\ref{consF}) is a consequence  of
(\ref{propF1}), (\ref{propF2}) and of $\FF$ being at each order in
$\la$ a finite sum of finite products of vectorfields} 
\eq
\label{consF}
\FF=1\otimes 1 + {\cal O}(\la)~.
\en
Property (\ref{propF1}) states that $\FF$ is a two cocycle, and it will turn out to be responsible for the associativity of the $\st$-products to be defined. Property 
(\ref{propF2}) is just a
normalization condition. From (\ref{consF}) it follows that 
${\cal F}$ can be formally inverted as a power series in $\lambda$. It also shows that the geometry we are going to construct has the nature of a deformation, i.e. in the $0$-th order in $\lambda$ we recover  the usual undeformed geometry.

Using the twist $\FF$ we now proceed to deform the commutative
geometry on $M$ into the twisted noncommutative one. 
The guiding principle is the observation that every time we have 
a linear map $X\otimes Y\rightarrow Z$, or a linear map 
$Z\rightarrow X\otimes Y$, 
where $X,Y,Z$ are vectorspaces, and where $\UU$ acts on $X, Y$ and $Z$,  
we can combine this map with an action of the twist. In this way
we obtain a deformed version of the initial linear  map. 
To preserve algebraic properties of the original maps very particular
actions of the twist $\FF$ have to be used.
 
As an example let $X=Y=Z=A$ where $A\equiv Fun(M)\equiv C^\infty(M)[[\la]]$ is the
algebra of smooth functions on $M$. The elements of $\UU$ act on $A$ by the natural 
extension of the Lie derivative. The Lie derivative on $Fun(M)$ associated with the vectorfield $v$ is defined as follows
\begin{equation}
{\cal L}_v(h) = v(h) \in A = Fun(M)~, \label{2.12}
\end{equation}
where $v\in \Xi$ and $h\in Fun(M)$. From equation (\ref{2.12}) follows
that the map
\begin{equation}
v\mapsto {\cal L}_v ,\label{2.13}
\end{equation}
satisfies
\begin{equation}
{\cal L}_{v'}{\cal L}_v(h) = v'(v(h)) \in Fun(M) \label{2.14}
\end{equation}
and therefore it is a Lie algebra homomorphism
\begin{equation}
[{\cal L}_{v'} ,{\cal L}_v ](h) = {\cal L}_{[v' ~ v]}(h)~. \label{2.15}
\end{equation}
This implies that we can
extend the Lie derivative associated with a vectorfield to a Lie
derivative associated with elements of $\UU$ by\footnote{Since
  $\ll_\xi$ is a differential operator, we have a map $\ll :
  \UU\rightarrow \Diff$ where $\Diff$ is the algebra of differential
  operators from $A$ to $A$. Notice that this map is neither surjective nor injective.}
 
\begin{equation}
{\cal L}_{\xi\zeta} = {\cal L}_\xi{\cal L}_\zeta~ .\label{2.16} 
\end{equation}
As in (\ref{2.12}) we frequently use the notation
\eq
\xi(h)=\ll_\xi(h)
\en
for the action of $\UU$ on $Fun(M)$.

The map we want to deform is the usual 
pointwise 
multiplication map between functions
\begin{eqnarray}
\mu: \quad  Fun(M) \otimes Fun(M) &\rightarrow & Fun(M)\nonumber\\
 f\otimes g &\mapsto & fg .\label{2.17}
\end{eqnarray}
To obtain $\mu_\star$ we first apply ${\cal F}^{-1}$ and then 
$\mu$ 
\begin{eqnarray}
\mu_\star: \quad  Fun(M) \otimes Fun(M) &{\stackrel{\,\FF^{-1}}{{-\!\!\!\rightarrow}}}& Fun(M)\otimes Fun(M)
{\stackrel{\mu}{\rightarrow}} Fun(M)\nonumber\\
f\otimes g ~~&\mapsto& ~{\cal F}^{-1}(f\otimes g)~ \mapsto~ \mu\circ {\cal F}^{-1}(f\otimes g) ~.\label{2.18}
\end{eqnarray}
This product is the $\st$-product
\eq\label{starprodf}
f\st g\equiv\mu_\st(f,g):=\mu\circ \FF^{-1}(f\otimes g)~.
\en
We see that $\mu_\star = \mu\circ {\cal F}^{-1}$ is a bidifferential operator.

That the $\star$-product is associative follows from (\ref{propF1}), 
see the theorem in Section \ref{Theorem1} for the proof. 
This is only true because we have used 
${\cal F}^{-1}$ and not ${\cal F}$ in (\ref{2.18}). We also have
\begin{equation}
f\st 1 = f = 1\st f \label{2.19}
\end{equation}
as a consequence of the normalisation condition (\ref{propF2}). From
(\ref{consF})
 follows that 
\begin{equation}
f\st g = fg + {\cal O}(\lambda) ~.\label{2.20}
\end{equation}
We have thus deformed the commutative algebra of function $A\equiv Fun(M)$ 
into the noncommutative one
\eq
A_\st\equiv Fun_\st(M)~.
\en

We shall frequently use the notation (sum over $\al$ understood)
\eq\label{Fff}
\FF=\f^\al\otimes\f_\al~~~,~~~~\FF^{-1}=\of^\al\otimes\of_\al~,
\en
so that
\eq\label{fhfg}
f\st g:=\of^\al(f)\of_\al(g)~.
\en
The elements $ \f^\alpha, \f_\alpha, \of^\al,\of_\al$ live in  $\UU$. 

In order to get familiar with this notation we will rewrite equation
(\ref{propF1}) and its inverse, 
\eq\label{ifpppop}
((\Delta\otimes id)\FF^{-1}) 
\FF^{-1}_{12} =((id \otimes \Delta)\FF^{-1})\FF^{-1}_{23}~,
\en 
as well as (\ref{propF2}) and (\ref{consF}) using the notation (\ref{Fff}), explicitly
\begin{eqnarray}
\f^\beta \f^\alpha_{_1}\otimes \f_\beta \f^\alpha_{_2}\otimes \f_\alpha &=& 
\f^\alpha\otimes \f^\beta \f_{\alpha_1}\otimes \f_\beta
\f_{\alpha_2} ,\label{2.21}\\
\label{ass}
\of_{_1}^\al\of^\be\otimes \of_{_2}^\al\of_\be\otimes \of_\al&=&
\of^\al\otimes {\of_{\al_1}}\of^\be\otimes {\of_{\al_2}}\of_\be~,\\
\varepsilon(\f^\alpha)\f_\alpha &=& 1 =  \f^\alpha\varepsilon(\f_\alpha) , \label{2.23}\\
{\cal F} = \f^\alpha\otimes \f_\alpha &=& 1\otimes 1 + {\cal O}(\lambda) .\label{2.24}
\end{eqnarray}

\subsection{Examples of twists} 
\sk
\noi {\bf 1)} Consider the case $M=\mathbb{R}^n$ and the element
\eq
\FF=e^{-{i\over 2}\hbar\theta^{\mu\nu}{\partial\over \partial x^\mu}
\otimes{\partial\over \partial x^\nu}}
\en 
where $\theta^{\mu\nu}$ is an antisymmetric matrix of real numbers.
The inverse of $\FF$ is
$$\FF^{-1}=e^{{i\over 2}\hbar\theta^{\mu\nu}{\partial\over \partial x^\mu}
\otimes{\partial\over \partial x^\nu}}~.$$
Then we have
$$(\Delta\otimes id)\FF=e^{-{i\over 2}\hbar\theta^{\mu\nu}
({\partial\over \partial x^\mu}\otimes 1
\otimes{\partial\over \partial x^\nu}+
1\otimes{\partial\over \partial x^\mu}
\otimes{\partial\over \partial x^\nu})}$$
so that property (\ref{propF1}) follows: 
$$\FF_{12}(\Delta\otimes id)\FF=e^{-{i\over 2}\hbar\theta^{\mu\nu}
({\partial\over \partial x^\mu}\otimes 
{\partial\over \partial x^\nu}\otimes 1+
{\partial\over \partial x^\mu}\otimes 1
\otimes{\partial\over \partial x^\nu}+
1\otimes{\partial\over \partial x^\mu}
\otimes{\partial\over \partial x^\nu})}
=\FF_{23}(id\otimes \Delta)\FF\,.$$
Property (\ref{propF2}) trivially holds.
The $\st $-product that the twist $\FF$ induces on the algebra of
functions on $\mathbb{R}^n$ is the usual $\theta$-constant $\st $-product (Moyal-Weyl $\star$-product),
\eq\label{starprod}
(f\st g)(x)=
e^{{i\over 2}\hbar\theta^{\mu\nu}{\partial\over \partial x^\mu}
{\partial\over \partial y^\nu}}f(x)g(y)|_{y\rightarrow x}~.
\en

\noi{\bf  2)} More in general on a smooth manifold $M$ consider a set of mutually
commuting smooth vectorfields $\{X_a\}$, $a=1,2,...s$. 
These vectorfields are globally
defined on the manifold $M$ but can be zero outside a given open of
$M$.
Consider then 
\eq\label{dcvfrrg}
\FF=e^{{\hbar\sigma^{ab}X_a\otimes X_b}}
\en
where $\sigma^{ab}$ are arbitrary constants.
The proof that $\FF$ is a twist is the same as that of the first example.

In the case that $M$ is a Lie group (and more generally
a quantum group) deformations of the form  (\ref{dcvfrrg}) appeared in 
\cite{Reshetikhin}.
See also \cite{Jambor} where a few examples 
that reproduce known $q$-deformed spaces are 
explicitly presented. 
\sk
\noi {\bf 2a)} 
A star product that implements the quantum plane commutation relation
$xy=qyx$ ($q=e^{i\hbar}$) can be obtained via the twist 
\begin{equation}
\FF=e^{-{i\over 2}\hbar(x{\partial\over \partial x}\otimes y{\partial\over
\partial y}-y{\partial\over \partial y}\otimes x{\partial\over \partial x})}.
\end{equation}
Notice that the vectorfields $x{\partial\over \partial x}$ and
$y{\partial\over \partial y}$ vanish at the origin.
In the semiclassical limit we have a Poisson structure not a
symplectic one. 
\sk
\noi {\bf 2b)} Consider the sphere $S^2$ and the usual polar coordinates 
$0\leq\varphi < 2\pi$, $0\leq\vartheta\leq \pi$. Let $f(\vphi)$ and 
$l(\vartheta)$
be arbitrary smooth functions with support for example in 
$(-{\pi\over 4}, {\pi\over 4})$ and  $({\pi\over 8}, {3\pi\over 8})$ respectively. 
Then 
\eq
\FF=e^{\hbar f(\vphi){\partial\over \partial\vphi}\otimes
l(\vartheta){\partial\over \partial\vartheta}}
\en 
gives a well defined star product on the sphere. 
\sk
\noi{\bf  3)} Twists are not necessarily related to commuting
vectorfields. For example consider on a smooth manifold $M$ four
vectorfields $H, E,A,B$, that satisfy the Lie algebra relations
\eqa
& & [H,E]=2E~,~~ [H,A]=\al A~,~~[H,B]=\be B~,~~~~~~~~\al+\be=2~,\nn\\[.1cm]
& & [A,B]=E~,~~~~[E,A]=0~,~~~\;~[E,B]=0~.   
\ena
Then the element
\eq
\FF=e^{{1\over 2}H\otimes_{} {\rm{ln}}(1+\la E)} \:  e^{\la A\otimes B{1\over 1+\la E}}
\en
is a twist and gives a well defined $\st$-product on the algebra of
functions on $M$.
These twists are known as extended Jordanian deformations
\cite{Kulish1}. Jordanian deformations 
\cite{GER, OGIEV} are obtained setting $A=B=0$ (and keeping the relation $[H,E]=2E$).

\subsubsection{Deformed Superspace}

Consider the superspace $\mathbb{R}^{m|n}$ with coordinates 
$(x^{\mu},\theta^{\alpha})\equiv Z^{A}$ 
and partial derivatives $(\partial_{\mu},\partial_{\alpha})\equiv\partial_{A}$
that satisfy the following (anti-)commutation relations\begin{eqnarray*}
[Z^{A},Z^{B}]_{\pm}=0 & , &
\partial_{A}Z^{B}=\delta_{A}^{B}~.\end{eqnarray*}
A generic derivation is of the form $\chi=f^A(Z)\partial_A$, where
$f^A(Z)$ are functions on superspace.
Consider a set $\{\chi_{a},\chi_{\epsi}\}\equiv\{\chi_{I}\}$ 
of even derivations $\chi_a$ and of odd derivations
$\chi_\epsi$ that are mutually (anti-)commuting, 
\eq \label{super}
[\chi_{I},\chi_{J}]_{\pm}=0~;
\en
for instance one can consider the derivativations
$\{\chi_{I}\}=\{\partial_{\mu},\partial_{\alpha}\}$, or
the derivations $\{\chi_{I}\}=\{{\partial\over\partial x^1},
\theta^1{\partial\over\partial\theta^1},
\theta^2{\partial\over\partial\theta^2},{\partial\over\partial\theta^3},\theta^4{\partial\over\partial x^2}\}$
(if $m\geq 2$ and $n\geq 4$).%

The universal enveloping superalgebra of the Lie
superalgebra (\ref{super}) is as usual the algebra $\mathcal U$ 
over $\mathbb{C}$  generated by the elements $\chi_I$ modulo the 
relations (\ref{super}).
The algebra $\mathcal U$ becomes a Hopf superalgebra
by defining on the generators the following grade preserving coproduct and antipode, and the following counit:
\begin{eqnarray*}
\Delta(\chi_{I}):=\chi_{I}\otimes1+1\otimes\chi_{I}~, ~~~ S(\chi_{I}):=-\chi_{I}~,~~~  \varepsilon(\chi_{I}):=0~,\end{eqnarray*}
where the tensorproduct $\otimes$ is over $\mathbb{C}$.
The multiplication in $\mathcal
U\otimes \mathcal U$ is defined as follows for
homogeneous elements $\xi,\zeta,\xi',\zeta'\in\mathcal{U}$ (of even or
odd degree
$|\xi|,|\zeta|,|\xi'|,|\zeta'|$ respectively):
\begin{equation}\label{graprouu}
(\xi\otimes \zeta)(\xi'\otimes \zeta')=(-1)^{|\zeta||\xi'|}\xi\xi'\otimes \zeta\zeta'.
\end{equation}
The antipode is extended to all elements of $\mathcal{U}$ by requiring it to 
be linear and graded antimultiplicative; the coproduct is linear and 
multiplicative (the grading being already present in (\ref{graprouu}){});
the counit is linear and multiplicative: 
\eq
\D(\xi\zeta)=\D(\xi)\D(\eta)~~~,~~~~~
S(\xi\zeta)=(-1)^{|\zeta||\xi|}S(\zeta)S(\xi)~~~,~~~~
\epsi(\xi\zeta)=\epsi(\xi)\epsi(\zeta)~.
\en
We refer to \cite{Frabetti} for a concise treatment of Hopf superalgebras. 

Consider the even element in  $\mathcal U[[\la]]\otimes \mathcal
U[[\la]]$ given by
\eq\label{susytwist}
\mathcal{F}:=e^{\la\sigma^{IJ}\chi_I\otimes\chi_J}=
e^{\la\sigma^{aa'}\chi_a\otimes\chi_{a'}+\la\sigma^{\epsi\epsi'}
\chi_\epsi\otimes\chi_{\epsi'}}~,
\en
where $\{\sigma_{IJ}\}\equiv\{\sigma_{aa'},\sigma_{\epsi\epsi'}\}$
are arbitrary constants ($\mathbb{C}$-numbers).
In order to check that
$\mathcal{F}$ is a twist as defined in 
Definition \ref{twistdeff} we observe that
$
\FF_{12}=e^{\la\sigma^{IJ}\chi_I\otimes\chi_J}\otimes 1=
e^{\la\sigma^{IJ}\chi_I\otimes\chi_J\otimes 1}
$, and that 
\eq
{(\Delta\otimes id)}\FF=
e^{\la\sigma^{IJ}(\chi_I\otimes 1+1\otimes\chi_I)\otimes\chi_J}~.
\en
This last relation holds because $\D\otimes id\,:\,\mathcal U\otimes \mathcal U \rightarrow \mathcal U\otimes \mathcal U\otimes \mathcal U$ is multiplicative 
(the product in $\mathcal U\otimes \mathcal U\otimes \mathcal U$ is given by
($\xi\otimes \zeta\otimes \eta)(\xi'\otimes \zeta'\otimes\eta')=(-1)^{|\zeta+
\eta||\xi'|+|\eta||\zeta'|}\xi\xi'\otimes \zeta\zeta'\otimes\eta\eta'_{\,}$).
Finally 
\eq\FF_{12}{(\Delta\otimes id)}\FF=
e^{\la\sigma^{IJ}\chi_I\otimes\chi_J\otimes 1 \,+\,
\la\sigma^{IJ}(\chi_I\otimes 1+1\otimes\chi_I)\otimes\chi_J}
\en
because the arguments of the exponentials are even elements of
$\mathcal U\otimes \mathcal U\otimes \mathcal U$ whose commutator vanishes.  
One similarly computes $\FF_{23}(id\otimes \FF)$.

An associative $\star$-product on superspace is then defined by
\begin{eqnarray}
g\star h & := & \mu\circ\mathcal{F}^{-1}(g\otimes h)\nonumber \\
 & = &
 (-1)^{|\of_{\alpha}||g|}\,\of^{\alpha}(g)\,\of_{\alpha}(h)~.\label{eq: definition of superstar}\end{eqnarray}
Associativity depends only on property (\ref{ass}) and not on the specific
example of twists (\ref{susytwist}). Associativity is explicitly proven in 
Appendix \ref{susyass}.

As particular cases of this construction we obtain the 
non anti-commutative superspaces considered in \cite{1}. 
For twists on superspace see also \cite{Ihl} and references therein.

\subsection{The deformed Hopf algebra $\UU^{\cal F}$}\label{S2.4}

Another deformation via the action of ${\cal F}$ leads to a new Hopf 
algebra 
\begin{equation}
(\UU ^{\cal F},\cdot, \Delta^{\cal F}, S^{\cal F}, \varepsilon ^{\cal
  F}) = (\UU ,\cdot, \Delta^{\cal F}, S^{\cal F}, \varepsilon). \label{2.4.1}
\end{equation}
As algebras $\UU^\FF =\UU$ and they also have the same counit 
$\varepsilon^\FF=\varepsi$. 
The new coproduct $\Delta^{\cal F}$ is given by
\begin{eqnarray}
\Delta^{\cal F}: \quad && \UU^\FF=\UU {\stackrel{\Delta}{\longrightarrow}} 
\UU \otimes \UU\,
{\stackrel{Conj_\FF}{-\!\!\!-\!\!\!-\!\!\!\longrightarrow}}~ \UU\otimes\UU=\UU^{\FF}\otimes \UU^{\cal F} \nonumber\\
&&~~~~~\,\,~\xi ~\mapsto~~ ~\Delta(\xi)~ ~\mapsto~~ \Delta^{\cal F}(\xi) = {\cal F}\Delta(\xi){\cal F}^{-1} .\label{2.4.1bis}
\end{eqnarray}
We deform the antipode, a map from $\UU$ to $\UU$, using
an invertible element $\chi$ of $\UU$ defined as follows\footnote{See Appendix \ref{Hopf F} for a proof that $\chi\chi^{-1}=\chi^{-1}\chi=1$.} 
\begin{equation}
\chi := \f^\al S(\f_\al) \quad,\quad \chi^{-1} = S(\of^\al) \of_\al~ .\label{2.4.2}
\end{equation}
The definition of the new antipode is
\begin{equation}
S^\FF(\xi)=\chi S(\xi)\chi^{-1} .\label{2.4.3}
\end{equation}
We follow the same steps as in Subsection (\ref{SEct2.1}) to show that 
$\UU^\FF = (\UU ^{\cal F},\cdot, \Delta^{\cal F}, S^{\cal F}, \varepsilon)$
is a Hopf algebra.

That $\Delta^{\cal F}$ and $\varepsilon$ are algebra homomorphisms and that $S^{\cal F}$ is an antialgebra homomorphism follows immediately from the definition
\begin{eqnarray}
\Delta^{\cal F} (\xi\zeta) &=& \Delta^{\cal F}(\xi)\Delta^{\cal F}(\zeta)~ ,\nonumber\\
\varepsilon^{\cal F} (\xi\zeta) &=& \varepsilon^{\cal F}(\xi)\varepsilon^{\cal F}(\zeta)~ ,\nonumber\\
S^{\cal F}(\xi\zeta) &=& S^{\cal F}(\zeta)S^{\cal F}(\xi)~ .\label{2.4.4}
\end{eqnarray}
We have now to show that $\Delta^{\cal F}$ and $S^{\cal F}$ fulfill
the additional conditions (\ref{2.6}), and therefore that 
$(\UU ^{\cal F},\cdot, \Delta^{\cal F}, S^{\cal F}, \varepsilon)$ 
is a Hopf algebra. This is done in
Appendix \ref{Hopf F}.

The new Hopf algebra $\UU^\FF$ is triangular, i.e.,
there exists an invertible element $\RR\in \UU^\FF\otimes \UU^\FF$ 
(called universal $\RR$-matrix) such that, for all $\xi\in \UU$,
\eqa\label{12}
{\Delta^\FF}^{\,op}(\xi)&=&\RR\Delta^\FF(\xi)\RR^{-1}\\
(\Delta^\FF\otimes id)\RR=\RR_{13}\RR_{23}~,&~&~~
(id\otimes \Delta^\FF)\RR=\RR_{13}\RR_{12}~,\label{13}\\
\RR_{21}&=&\RR^{-1}~\label{14}~,
\ena
where $\RR_{21}=\sigma (\RR)\in \UU^\FF\otimes\UU^\FF$,
with $\sigma $ the flip map, $\sigma(\xi\otimes\zeta)=\zeta\otimes\xi$.
The two equations in (\ref{13}) take value in $\UU\otimes\UU\otimes\UU$, and 
$\RR_{12}=\RR\otimes 1, \RR_{23}=1\otimes \RR$, while 
$\RR_{13}\in\UU\otimes\UU\otimes\UU$
has the unit $1$ in the middle factor.
 Defining
\eq
\RR:=\FF_{21}\FF^{-1}~
\en
it can be shown that 
equations (\ref{12}), (\ref{13}), (\ref{14}) are
fulfilled. The
cocycle condition of $\FF$ was in this context only needed to prove (\ref{13}) 
\footnote{We refer to \cite{Majidbook} (p. 56), see also \cite{Chari-Pressley} (p.130),
for a proof of (\ref{13}) and for an 
introduction to twists and their relations 
to Hopf algebra deformations.}.
In the sequel we use the notation 
\eq
\RR=\R^\al\otimes\R_\al~~~,~~~~~~\RR^{-1}=\oR^\al\otimes\oR_\al~.
\en
Using the notation introduced in (\ref{Fff}) we obtain
\begin{equation}
\RR=\R^\al\otimes\R_\al=\f_\al\of^\be \otimes \f^\al\of_\be~~~, \qquad \RR^{-1}=\oR^\al\otimes\oR_\al = \f^\al\of_\be \otimes \f_\al \of^\be~.\label{2.4.2bis}
\end{equation}


\section{Representations}\label{Sec3}

\subsection{Module Algebras}\label{Theorem1}

Having a Hopf algebra, its modules are certainly of interest in
physics and mathematics. They are the representations of the Hopf
algebra. Here we show that to a module algebra $\AA$ of the Hopf 
algebra $\UU$
there corresponds a module algebra $\AA_\st$ of the deformed Hopf 
algebra
$\UU^{\cal F}$. 

A module algebra $\AA$ is  a module $\AA$ 
on which $\UU$ acts which in addition has an algebra structure that 
is compatible with the action of $\UU$,
for all $\xi\in \UU$ and $ a,b\in \AA$,
$$\xi(ab)=\mu\circ\Delta(\xi)(a\otimes b)=\xi_1(a)\xi_2(b)~~~,~~~~
\,\xi(1)=\varepsilon(\xi)1\,. $$
(where $1$ is the unit in $\AA$). 

\sk
We recall a basic theorem
concerning representations of twisted Hopf algebras.
Given a twist $\FF\in\UU\otimes\UU$, 
we can construct a deformed algebra $\AA_\st$.
The algebra $\AA_\st$ has the same vector space 
structure as $\AA$ and the action of $\UU^\FF$ on $\AA_\st$ is the
action of $\UU$ on $\AA$.
The product in $\AA_\st $ is defined by
\eq
a\st  b=\mu\circ \FF^{-1}(a\otimes b)=\of^\al(a)\of_\al(b)~,
\en
in accordance with formula (\ref{starprodf}).
Compatibility between the action of $\UU^\FF$ and the product in
$\AA_\st$  demands
\eq\label{eq46p}
\xi(a\st b)=\xi_{1_\FF}(a)\st\xi_{2_\FF}(b)~,
\en
where we used the notation $\Delta^\FF(\xi)=\xi_{1_\FF}\otimes \xi_{2_\FF}$.

In order to prove associativity of the new product we
use (\ref{ass}) and compute:
\eqa
(a\st b)\st c &=&\of^\al(\of^\be(a)\of_\be(b))\of_\al(c)=
(\of_{_1}^\al\of^\be)(a) (\of_{_2}^\al\of_\be)(b) \of_\al(c)\nn
=
\of^\al(a) ({\of_{\al_1}}\of^\be)(b)
({\of_{\al_2}}\of_\be)(c)\nn \\
&=&
\of^\al(a)\of_\al(\of^\be(b)\of_\be(c))=a\st (b\st c)\nn~.
\ena
We still have to prove (\ref{eq46p}):
$$\xi(a\st b)=\xi(\mu\circ \FF^{-1}(a\otimes b))=
\mu\circ \D(\xi)\circ\FF^{-1}(a\otimes
b)=\mu\circ\FF^{-1}\circ\D^\FF(\xi)(a\otimes
b)
=\xi_{1_\FF}(a)\st\xi_{2_\FF}(b)~.
$$
Notice also that if $\AA$ has a unit element $1$, then
$1\st a=a\st1$ follows from the
normalization condition property (\ref{propF2}) 
of the twist $\FF$.

\subsection{Examples of Module algebras}\label{Examples_Modalgs}
We now apply this construction
to the $\UU$-module algebras $A$ and $\UU$. 
In both cases the action of $\UU$ on the corresponding module algebra is given by the Lie derivative. 


\sk
\noi{\bf{Algebra of noncommutative functions $A_\st$}}~~\\
We start with the $\UU$-module algebra of functions $\AA=A=Fun(M)$, and we
obtain the algebra $A_\st \equiv Fun_\st(M)$
with the $\star$-product already introduced in (\ref{starprodf}). 
The algebra $A_\st$, according to Section \ref{Theorem1} is a left $\UU^\FF$-module
algebra. In particular, vectorfields $u\in \Xi \subset \UU^{\FF}$ 
act according to the deformed Leibniz rule
\eq
u(h\st g)=u_{1_\FF}(h)\st  u_{2_\FF}(g),
\en 
where 
\eq \label{Leibniz_u}
\Delta^{\FF}(u)= u_{1_{\FF}} \otimes u_{2_{\FF}} = \f^\al u \,\of^\be \otimes \f_\al \of_\be + \f^\al  \of^\be \otimes \f_\al u \,\of_\be~.  
\en     
\sk

\noi {\bf The algebra $\UU_\st $}~\\
We next consider the case $\AA=\UU$. This is a module algebra with respect to the Hopf algebra $\UU$. The action of  $\UU$ on $\UU$ is given by the 
extended Lie derivative (adjoint action): the action of $\ll_u$ on $v$
is just the Lie bracket $\ll_u(v)=[u\,v]$; the action of $\UU$ on
$\Xi$ is obtained from the action of vectorfields by definining 
$\ll_{\xi\zeta}=\ll_\xi\ll_\zeta$
(where composition of the actions $\ll_\xi$ and $\ll_\zeta$
is understood); finally the action of $\UU$ on $\UU$ is obtained from 
the known Leibniz rule $\ll_u(vz)=\ll_u(v)z+v\ll_u(z)$, that implies 
$\ll_{\xi}(\zeta\eta)=\ll_{\xi_1}(\zeta)\ll_{\xi_2}(\eta)$.
\sk
The deformed algebra $\UU_\st$ equals $\UU$ as a vectorspace, but it has the deformed product 
\eqa 
\st :~ \UU\otimes \UU & \rightarrow & \UU \nonumber \\
(\xi,\zeta) & \mapsto & \xi \st \zeta := \of^\al(\xi)\,\of_\al(\zeta) \label{starinUU}
\ena
where $\of^\al(\xi)$, (and  $\of_\al(\zeta)\,$) is 
another notation for the Lie derivative 
$\ll_{\of^\al}(\xi)$,  (and $\ll_{\of_\al}(\zeta)\,$). 
The Hopf algebra $\UU^\FF$ acts on $\UUs$, and compatibility with the
$\st$-product of $\UUs$ is 
\eq
\xi(\zeta \star \eta)= \xi_{1_\FF}(\zeta)\star\xi_{2_\FF}(\eta)\,.
\en  

This way we have obtained from the theorem 
in Section 3.1 the algebra $\UU_\st$. 
We will show in Section 3.3 that it is a Hopf algebra. 

In $\UU_\st$ we consider the deformed commutator
of the vectorfields  $u,v\in\Xi$,
\eq
[u,v]_\st:=u\star v-\oR^\alpha(v)\star\oR_\alpha(u)~.
\en
This commutator closes in $\Xi$:
\eqa
u\star v-\oR^\alpha(v)\star\oR_\alpha(u) & = & \of^\ga(u) \of_\ga(v) - \of^\ga(\oR_\alpha(v)) \of_\ga(\oR^\alpha(u)) \nonumber \\
& = & \of^\ga(u) \of_\ga(v) - \of^\ga \f^\al \of_\be(v) \of_\ga \f_\al \of^\be(u) \nonumber \\
& = &  \of^\ga(u) \of_\ga(v) - \of_\ga(v) \of^\ga(u) \nonumber \\
& = & [\of^\ga(u)\,,\of_\ga(v)]~, \nonumber   
\ena
(the first line uses the definition of the $\st$-product, the second line the definition of the $\RR$-matrix, $\RR^{-1} = \oR^\al \otimes \oR_\al = 
\f^\al \of_\be \otimes \f_\al \of^\be$
 as introduced in Section \ref{S2.4}. The third line uses $\FF^{-1}\FF=1$).
The last term is a a sum (over $\ga$) of undeformed commutators between the vectorfields $\of^\ga(u)$ and $\of_\ga(v)$, 
and therefore $[u\,,v]_\st\in\Xi$.
 
We denote by $\Xi_\star$ the linear space
of vectorfields $\Xi$ equipped with the multiplication
\eqa
[~,~ ]_\st\,: \quad \Xi\times\Xi &\to& \Xi \nonumber\\
(u,v) &\mapsto& [u\,,v]_\st\, .\label{2.1st}
\ena
this way $\Xis$ becomes a deformed Lie algebra.
The elements of $\Xis$ we call $\st $-vectorfields.
It is easy to see that the bracket  $[~,~]_\st $ 
has the $\st$-antisymmetry property
\eq\label{sigmaantysymme}
[u\,,v]_\st =-[\oR^\al(v)\,,\, \oR_\al(u)]_\st~ .
\en
This can be shown as follows
$$[u\,,v]_\st =[\of^\al(u)\,,\of_\al(v)]=-[\of_\al(v)
\,,\of^\al(u)]=
-[\oR^\al(v)\,, \oR_\al(u)]_\st~. $$
We recall that $\RR^{-1}=\oR^\al\otimes\oR_\al
=\FF\FF_{21}^{-1}\in \UU\otimes\UU\,.$

A $\st$-Jacoby identity can be proven as well
\eq
[u \,,[v\,,z]_\st ]_\st =[[u\,,v]_\st \,,z]_\st  + [\oR^\al(v)\,,\, [\oR_\al(u)\,,\,z]_\st ]_\st ~.
\en
A direct proof of the
$\st $-Jacobi identity  can be found in Appendix \ref{AJacoby}.

\sk
Finally we notice that  any sum of
products of vectorfields in $\UU$ can be rewritten as sum of
$\st$-products of vectorfields via the formula $u\,v=\f^\al(u)\st\f_\al(v)$, and therefore
$\st$-vectorfields generate the algebra. 

Indeed we have  proven, see \cite{NG}, 
that $\UU_\st$ is the universal enveloping
algebra of $\Xis$.
\subsection{$\UU_\star$ is a Hopf algebra}
We have seen that $\UU$ can be equipped with
the usual Hopf algebra structure
$(\UU,\cdot,\Delta,S,\epsi)$ or with the twisted Hopf algebra 
$(\UU^\FF,\cdot,\Delta^\FF,S^\FF,\epsi)$ or with a new product
$\UU_\st=(\UU,\st)$. It turns out that $\UU_\st$ has
also a natural Hopf algebra structure,
\eq
(\UU_\st,\st,\Delta_\st,S_\st,\epsi_\st)~.
\en
We describe it by giving the
coproduct, the inverse of the antipode 
and the counit on the generators $u$ of $\UU_\st$:
\eq\label{coproductu}
\D_\st (u)=u\otimes 1+ X_{\oR^\al}\otimes {\oR_{\al}}(u)
\en
\eq
S^{-1}_\st(u)=-\oR^\al(u)\st X_{\oR_\al}~.
\en
\eq
\epsi_\st(u)=\epsi(u)=0~,\label{epsist}
\en
where, for all $\xi\in\UU$,  
$X_\xi=\of^\al\xi\chi S^{-1}(\of_\al)$. The map $X: \UU\rightarrow\UU$ 
is invertible and it can be shown \cite{Majid-Gurevich}, 
that its inverse $X^{-1}$ is
\eq
X^{-1}=\of^\al(\xi)\of_\al=:D(\xi)~.
\en
In principle one could directly check that
(\ref{coproductu})-(\ref{epsist}) define a bona fide Hopf algebra.
Another way \cite{NG} is to show that the Hopf algebra $\UU_\st$ is 
isomorphic to the Hopf algebra $\UU^\FF$. The isomorphism is 
given by the map
$D\,$: 
\eqa \label{D alg-homo}
&&D(\xi\st \zeta)=D(\xi)D(\zeta)~,\\
&&\D_\st =(D^{-1}\otimes D^{-1})\circ \D^\FF\circ D~,\label{DST}\\
&&S_\st =D^{-1}\circ S^\FF \circ D~.
\ena
In particular, since $\UU^\FF$ is a triangular Hopf algebra, also
$\UU_\st$
is a triangular Hopf algebra. Its $R$-matrix is 
\eq
\RR_\st =(D^{-1}\otimes D^{-1})(\RR)~~,~~~
\RR_\st =\R_\st ^\al\otimes \R_{\st \,\al}=X_{\R^\al}\otimes
X_{\R_\al}~.
\en 
Explicitly  we have 
\eqa
{\Delta_\st }^{op}(\xi)&=&\RR_\st \st \Delta_\st (\xi)\st \RR_\st
^{-1}\label{copopst}\\
(\Delta_\st \otimes id)\RR_\st =\RR_{\st \,13}\st \RR_{\st\,23}~,&&~~
(id\otimes \Delta_\st )\RR_\st =\RR_{\st \,13}\st \RR_{\st \,12}~,\\
\RR_{\st \,21}&=&\RR_\st ^{-1}\, ,\label{triangofRst}
\ena
where $\RR_\st ^{-1}$ is the $\st $-inverse of $\RR_\st $, 
i.e., $\RR_\st ^{-1}\st \RR_\st  = \RR_\st \st \RR_\st ^{-1}=1\otimes 1$.

\sk
Summarizing we have encountered the Hopf algebras 
$$(\UU,\cdot,\Delta,S,\epsi)~~,~~~
(\UU^\FF,\cdot,\Delta^\FF,S^\FF,\epsi)~~,~~~
(\UU_\st ,\st ,\Delta_\st ,S_\st ,\epsi)~.~$$
The first is cocommutative, the second is triangular and is 
obtained twisting the first, the third is triangular and isomorphic 
to the second. The remarkable fact about $\UU_\st$ is 
the Leibniz rule for vectorfields
(\ref{coproductu}). We have that $\oR_\al(u)$ is again a
vectorfield so that
\eq\label{woro}
\Delta_\st (\Xis) \subset \Xis \otimes 1 + \UU_\st\otimes\Xis~.
\en
This is a fundamental property for the construction of a deformed 
differential calculus \`a la Woronowicz \cite{Woronowicz}. Note that the 
coproduct $\D^\FF (u)$ does not have this property, 
as can be seen explicitly from (\ref{Leibniz_u}). 
It is interesting to note that a Hopf algebra with  comultiplication 
structure (\ref{Leibniz_u}) is  isomorphic to a Hopf algebra 
with comultiplication structure  (\ref{woro}).
In order to establish a gravity theory which is invariant with
respect to deformed infinitesimal diffeomorphisms 
we will consider module algebras with respect to
$\UU_\star$ and not with respect to $\UU^\FF$.

\section{Representations of deformed infinitesimal diffeomorphisms}
In Section \ref{Sec3} we have constructed the Hopf algebra $\UU_\st$.
Since $\UU_\star$ and $\UU^\FF$ are isomorphic as Hopf algebras, any
$\UU^\FF$-module has automatically a $\UU_\st$-module structure. 
In particular $A_\st$ and $\UU_\st$ are also 
$\UU_\st$-module algebras.

The action $\lls$ of $\UU_\st$ on $\AAs$  is given 
by combining the usual action (Lie derivative $\ll$) with 
the twist $\FF$
\eq
\ll^\st_\xi(h):=\ll_{\of^\al(\xi)}(\ll_{\of_\al}(h))~,
\en
or equivalently, recalling that $D(\xi)=\of^\al(\xi)\of_\al$, we see that
\eq\label{staraction}
\ll^\st _\xi:=\ll_{D(\xi)}~.
\en
Similarly for the action of $\UU_\st$ on $\UU_\st$, that we also denote
by $\lls$,
\eq
\ll^\st_\xi(\zeta):=\ll_{\of^\al(\xi)}(\ll_{\of_\al}(\zeta))
=\of^\al(\xi)(\of_\al(\zeta))~.
\en
It is easy to see that these actions are well defined:
$\lls_\xi\circ\lls_\zeta=\lls_{\xi\st\zeta}$,
for example we find\footnote{In \cite{Dimitrijevic}, \cite{G1},
\cite{Meyer}, we have $\theta^{\mu\nu}$-constant noncommutativity
and differential operators $X^\st_u$ that satisfy
$X^\st_u\st X^\st_v=X^\st_{uv}$, the relation between
$X^\st_u$ and $\lls_u$ (for the $\theta^{\mu\nu}$-constant case) 
is $(X^\st_u\st g)=u(g)=\lls_{X_u}(g)$.
}
\eq\label{comp0st}
\lls_\zeta(\lls_\xi(h))=\lls_\zeta((D\xi)(h))=
(D\xi)(D\zeta)(h)=D(\xi\st\zeta)(h)=\lls_{\xi\st\zeta}(h)
\en
where we used (\ref{D alg-homo}). 
Compatibility with the $\st$-product in $\AAs$ 
is also easily proven,
\eqa\label{comp1st}
\lls_\xi(h\star g)&=&\ll_{D\xi}(h\st g)=(D\xi)(h\st g)=
(D\xi)_{1_\FF}(h)\st(D\xi)_{2_\FF}(g)=
D(\xi_{\1s})(h)\st D(\xi_{\2s})(g)\nn\\
&=&\lls_{\xi_\1s}(h)\st\lls_{\xi_\2s}(g)
\ena
where we used (\ref{DST}). One proceeds similarly for the action $\lls$
of $\UU_\st$ on $\UU_\st$. The proofs that this action is well defined
and that it is compatible with the $\st$-product in $\UU_\st$, are 
exactly the same as in (\ref{comp0st}) and (\ref{comp1st}), just
substitute $h,g\in\AAs$ with $\zeta,\eta\in \UU_\st$.
We here notice in particular that the $\st$-Lie derivative
of a vectorfield on a vectorfield gives the $\st$-Lie bracket,
\eq
\lls_u(v)=[u\,,v]_\st~.
\en
Morover it can be shown  that the $\st$-Lie derivative of $\UU_\st$
on $\UU_\st$ equals the $\st$-adjoint action, 
$\lls_\xi(\zeta)=ad^\st_\xi(\zeta)\equiv 
\xi_\1s\st\zeta\st S_\st(\xi_\2s)$.
In particular the $\st$-commutator $[u\,,v]_\st$ is just the 
$\st$-adjoint action of $u$ on $v$.

\subsection{Tensorfields 
}
Our main interest in this subsection is the deformed 
algebra of tensorfields.
We recall that tensorfields on a smooth manifold 
can be described as elements in\footnote{We assume for simplicity that 
$\Om\otimes\ldots\Om\otimes\Xi\otimes\ldots\Xi
\cong\Gamma(T^\ste M\otimes\ldots TM\otimes
TM\otimes\ldots TM)$.
That this is always the case for a smooth manifold $M$ (see for example 
\cite{Gracia-Bondia}, Prop. 2.6.) follows from the existence of a
finite covering of $M$ 
that trivializes the tangent bundle $TM$ and the cotangent bundle $T^\ste M$, see for example 
\cite{Conlon},  Thm. 7.5.16.}
\eq
\Om\otimes\Om\otimes\ldots\Om\otimes\Xi\otimes\Xi\otimes\ldots\Xi
\en
where $\otimes$ here stands for $\otimes_A$.
Functions are in particular type $(0,0)$-tensorfields and 
the tensorproduct between a function and another tensorfield is as usual not 
explicitly written.
The tensorproduct is an 
associative product. This in particular implies
$\tau\otimes h\tau'=\tau h\otimes\tau'$ and
$h(\tau\otimes\tau')=(h\tau)\otimes\tau'$.
Tensorfields are a $\UU$ module, the action of $\UU$ on $\TT$ 
is obtained via the Lie derivative on tensorfields, 
that extends to a map
$\ll: \UU\otimes \TT\rightarrow \TT\,$. For example
$\ll_{uv}(\tau)= \ll_u(\ll_v(\tau))$.

By using the theorem in Section \ref{Theorem1} and by setting $\AA=\TT$ where
$\TT$ is the commutative algebra of tensorfieds, we obtain a 
deformed tensor algebra $\TT_\st$ with associative $\st$-tensor product 
\eq\label{defofthetensprodst}
\tau\otimes_\st\tau':=\of^\al(\tau)\otimes \of_\al(\tau')~.
\en 
It follows that in $\TT_\st$ we have in particular
\eqa
\tau\otimes_\st h\st\tau'&=&\tau\st h\otimes_\st\tau'~,\\
\label{assocstotimes}
~~~h\st(\tau\otimes_\st\tau')&=&(h\st\tau)\otimes_\st\tau'~.
\ena \label{assocstotimes123}
The $\st$-product between a function and a tensor is noncommutative
\eq\label{sttprod}
\tau\st h
=\ll_{\of^\al}(\tau)
\,\ll_{\of_\al}(h)=\ll_{\of_\al}(h)\,\ll_{\of^\al}(\tau)
=\ll_{\oR^\al}(h)\st \ll_{\oR_\al}(\tau)=
{\oR^\al}(h)\st \oR_{\al}(\tau)~.
\en
\sk


We now consider the construction performed at the 
beginning of this section, but with $\TT_\st$ instead of 
$\AAs$ (or $\UU_\st$) and obtain that $\TT_\st$ is a $\UU_\st$-module 
algebra.
The action of $\UUs$ on $\TT_\st$ is given by the $\st$-Lie 
derivative 
\eq\label{actionof llsxitau}
\ll^\st_\xi(\tau):=\ll_{D\xi}(\tau)=\of^\al(\xi)(\of_\al(\tau))~.
\en
Compatibility with the $\st$-product in $\TT_\st$ 
is proven as in (\ref{comp1st})
\eq
\lls_\xi(\tau\star \tau')=\lls_{\xi_\1s}(\tau)\st\lls_{\xi_\2s}(\tau')\nn~.
\en
In particular the $\st$-Lie derivative along vectorfields satisfies the deformed Leibniz rule
\eq
\lls_u(h\st g)=\lls_u(h)\st g + \oR^\al(h)\st \lls_{\oR_\al(u)}(g)~.
\en
in accordance with the coproduct formula (\ref{coproductu}).


\subsubsection{Vectorfields $\Xis$ are an $\AAs$-bimodule }
From the definition of the  product of tensorfields 
(\ref{defofthetensprodst}), considering  
functions and vectorfields as particular tensors, we see that 
we can $\st$-multiply functions with vectorfields from 
the left and from the right. Because of associativity of the
tensorprduct we see that the space of 
vectorfields $\Xis$ is an $\AAs$-bimodule. 
In the commutative case left and right action of functions 
on 
vectorfields coincide, 
$uh=hu\,$\footnote{Here $uh$ is just the
vectorfield  that on a function $g$ gives $(uh)(g):=u(g)h$.
This notation should not be confused with the operator 
notation $u\circ h=u(h)+hu$.}.
In the noncommutative case the left and right $\AAs$-actions on $\Xis$
are not the same, but are related as in (\ref{sttprod}).

\sk

\noi{\bf Local coordinates description of vectorfields}\\
In a coordinate neighborhood $U$ with coodinates $x^\mu$
any vectorfield $v$ can be expressed in the $\partial_\mu$ basis as
$v=v^\mu\partial_\mu$. We have a similar situation in the
noncommutative case. 
\sk
\noi {\bf Lemma 1}
In a coordinate neighborhood $U$ with coordinates $x^\mu$
every vectorfield $v$ can be uniquely written as
\eq
v=v_\st^\mu\st\partial_\mu~.\label{decomposition}
\en
where $v^\mu_\st$ are functions on $U$.
\begin{proof}
We know that
$v$ can be uniquely written as
$v=v^\mu\partial_\mu$. In order to prove  decomposition
(\ref{decomposition}) we show that the equation 
\eq\label{condst}
v_\st^\mu\st\partial_\mu=v^\mu\partial_\mu
\en
uniquely determines order by order in $\la$ the coefficients 
$v^\mu_\st$ in terms of the $v^\mu$ ones.
First we 
expand $v^\mu$,
$v_\st^\mu$ and $\FF^{-1}$,
\eqa 
&&v^\mu=v_0^\mu+\la v_1^\mu+\la^2 v_2^\mu+\ldots~~~~,~~~~~~
v_\st^\mu=v_{\st 0}^\mu+\la v_{\st 1}^\mu+\la^2 v_{\st 2}^\mu+\ldots
\\[0.5em] &&\FF^{-1}=\of^\al\otimes\of_\al=1\otimes 1+\la\, 
\of^{\al_1}\otimes\of_{\al_1}+\la^2\,
\of^{\al_2}\otimes\of_{\al_2}+\ldots\label{expansionofF}
\ena
Then from (\ref{condst}) we have
\eq 
v_{\st 0}^\mu = v_0^\mu~~~,~~~~v_{\st 1}^\mu =
v_1^\mu-\of^{\al_1}(v^\rho)\of_{\al_1\,\rho}^{\,\mu}
\en
where $\of_{\al_1\,\rho}^{\,\mu}\partial_\mu=\of_{\al_1}(\partial_\rho)$. 
More in general at order $\la^i$ we have the equation
$v_{\st i}^\mu\partial_\mu+\sum_{j=1}^i\of^{\al_j}(v^\rho_{\st i-j})
\of^{\,\mu}_{\al_j\,\rho}\partial_\rho=v_i^\mu\partial_\mu$
that uniquely determines $v_{\st i}$ in terms of $\FF$, $v^\mu$ and
$v^\mu_{\st j}$ with $j<i$.
\end{proof}
Notice that this proof remains true if the local frame 
$\{\partial_\mu\}$ is replaced by a more general 
(not necessarily holonomic or $\la$ independent) frame $\{e_a\}$.
(Hint: $e_a=e_a^\mu\st\partial_\mu$, $\partial_\mu=e_\mu^a\st e_a$).

Along these lines one can define a change of reference frame, 
\eq\label{transfG}
\partial_\mu\rightarrow \partial'_\mu=L^\nu_{~\mu}\partial_\nu=
L^\nu_{\st\,\mu}\st\partial_\nu~.
\en 
This is a starting point in order to construct noncommutative
transition functions for the tangent bundle $TM$.

\subsubsection{1-Forms $\Omega_\st$}
From the tensorfield product definition 
(\ref{defofthetensprodst}), we see that
the space of 1-forms is an $\AAs$-bimodule.
The $\AAs$-bimodule structure explicitly reads, 
$\forall h\in \AAs, \omega\in \Omega_\st$,
\eq
\omega\st h=\lls_{\oR_\st^\al}(h)\st \lls_{\oR_{\st \al}}
(\omega)={\oR^\al}(h)\st \oR_{\al}(\om)~.
\en
The action of $\UU_\st$ on $\Oms$ is given in 
(\ref{actionof llsxitau}).
\sk
\noi{\bf Local coordinates description of 1-forms and of tensorfields}

\noi
As in the case of vectorfields we have 
that in a coordinate neighborhood $U$ with coordinates $x^\mu$
every 1-form $\omega$ can be uniquely written as
\eq\label{form85form}
\omega=\omega^\st_\mu\st dx^\mu
\en
with $\omega^\st_\mu$ functions on $U$, and where $\{dx^\mu\}$ is the usual 
dual frame of the vectorfields frame $\{\partial_\mu\}$.
We can now show
\sk
\noi{\bf Lemma 2}
In a coordinate neighborhood $U$ with coordinates $x^\mu$
every tensorfield $\tau^{p,q}$ can be uniquely  
written as 
\eq
\tau^{p,q}=\tau_{\st~\mu_1\ldots\mu_p}^{~\nu_1\ldots\nu_q}
\st dx^{\mu_1}\otimes_\st\ldots dx^{\mu_p}
\otimes_\st\partial_{\nu_1}\otimes_\st
\ldots\partial_{\nu_q}
\label{79}
\en
 where $\tau_{\st~\mu_1\ldots\mu_p}^{~\nu_1\ldots\nu_q}$ are functions
 on $U$.

\begin{proof}
Following the proof of Lemma 1 we have that 
$\tau^{p,q}$ can be uniquely written as
$\tau^{p,q}=\tau_\st^{p,q-1~\nu}\otimes_\st\partial_\nu$, 
where $\tau_\st^{p,q-1~\nu}$ is a type $(p,q-1)$ tensor. This
expression holds for any value of $q$ and therefore 
(using associativity of the $\otimes_\st$ product), $\tau^{p,q}$ can be uniquely written as
$\tau^{p,q}=
\tau_\st^{p,0~\nu_1\nu_2\ldots\nu_q}\otimes_\st
\partial_{\nu_1}\otimes_\st\ldots\partial_{\nu_q}
$. Similarly, like in formula (\ref{form85form}), we have 
that $\tau^{p,q}$ can be uniquely written as $\tau^{p,q}=
\tau_{\st~\mu_1}^{p-1,0~\nu_1\nu_2\ldots\nu_q}\otimes_\st dx^{\mu_1}
\otimes_\st
\partial_{\nu_1}\otimes_\st\ldots\partial_{\nu_q}\,.
$
 This
expression holds
for any value of $p$ and $q$ and therefore (using associativity 
of the $\otimes_\st$ product) we obtain expression (\ref{79})
and its uniqueness.
\end{proof}

\subsubsection{Exterior algebra of forms 
$\Omega^{\mbox{\boldmath $\cdot$}}_\st=\oplus_p\Omega^{p}_\st$}
\label{Sectexterioralgofform}
As another application of the theorem in
Section \ref{Theorem1} we consider the algebra 
of exterior forms $\Omega^{\mbox{\boldmath$\cdot$}}=\oplus_p\Omega^p$, 
and $\st$-deform the wedge product into the $\st$-wedge product,
\eq\label{formsfromthm}
\vartheta\wedge_\st\vartheta':=\of^\al(\vartheta)\wedge \of_\al(\vartheta')~.
\en 
We denote by $\Omega^{\mbox{\boldmath $\cdot$}}_\st$ 
the linear space of forms equipped with the wedge product  
$\wedge_\st$,
\begin{equation}\label{defOM}
\Omega^{\mbox{\boldmath $\cdot$}}_\star := (\Omega^{\mbox{\boldmath $\cdot$}},\wedge_\star) ~.
\end{equation} 
As in the commutative case it can be shown \cite{NG} that the linear space of exterior forms can be seen as the tensor subspace of totally 
$\st$-antisymmetric (contravariant) tensorfields. 
The properties of the $\st$-antisymmetrizator imply that there is a top form 
that has the same degree as in the undeformed case. 
This is in accordance with (\ref{formsfromthm}).
Explicitly the $\st$-antisymmetric 2-form 
$\omega\wedge_\st\omega'$ is defined by (cf. (\ref{transpstonforms})),
\eq
\omega\wedge_\st\omega':= \omega\otimes_\st\omega'
-\ll^\st_{\oR_\st^\al}(\omega')\otimes_\st\ll^\st_{\oR_{\st\al}}(\omega)~.
\en
It can also be shown \cite{Sitarz} 
that the usual exterior derivative 
$d:A\rightarrow \Om$ satisfies the Leibniz rule $d(h\st g)=dh\st g+h\st
dg $ and is therefore also the $\st $-exterior derivative. This is 
so because the exterior derivative commutes with the Lie derivative.
In the case that $A$
is a Hopf algebra, the fact that the exterior differential on $A_\st$ is
not deformed was shown in \cite{Majid-Oeckl}.

\subsection{$\st$-Pairing between 1-forms and vectorfields}\label{pairingformvect}
Following the general prescription outlined in Section \ref{THETWIST},
we define the $\st$-pairing between vectorfields and 1-forms
as $\le~,~\re_\st:=\le~,~\re\circ \FF^{-1}$. Explicitly, for all 
$\xi\in \Xis, \omega\in \Omega_\st, $
\eqa\label{lerest}
\le~,~\re_\st : \,
\Xi_\st\otimes_\bbC \Omega_\st &\rightarrow & A~,\\
(\xi,\omega)~&\mapsto &\le \xi,\omega\re_\st
:=\le\of^\al(\xi),\of_\al(\omega)\re~.
\ena
We leave it
to the reader to prove the following 
\sk
\noi{\bf Lemma 3 }
The pairing $\le~,~\re_\st$ is compatible with the 
$\st$-Lie derivative,
\eq\label{lieagainder}
\ll^\st_\xi(\le u,\omega\re_\st)=\le\ll^\st_{\xi_{1_\st}}(u),
\ll^\st_{\xi_{2_\st}}(\omega)\re_\st~,
\en
and satisfies the  $A_\st$-linearity properties  
\eq
\le h\st u,\omega\st k\re_\st=h\st\le u,\omega\re_\st\st k~,
\en
\eq
\le u, h\st\omega \re_\st=
\le u  \st h,
\omega\re_\st=
\ll^\st_{\oR_\st^\al}(h)\st\le \ll^\st_{\oR_{\st\al}}(u),\omega \re_\st~.
\en
so that $\le~,~\re_\st : \,
\Xi_\st\otimes_\st \Omega_\st \rightarrow  A~.$
\sk
In the commutative case we can consider locally a moving frame 
(or vielbein) $\{e_i\}$ and a dual frame of 1-forms $\omega^j$:
\eq
\le e_i,\om^j\re=\delta^j_i~,
\en 
in particular $\le \partial_\mu,dx^\nu\re=\delta_\mu^\nu$.
In the noncommutative case locally we also have a moving  
frame $\{\hat e_i\}$ and a dual frame of 1-forms $\omega^j$:
\eq
\le \hat e_i,\om^j\re_\st=\delta^j_i~.
\en 
We construct it in the following way: since $\le
e_i,\om^j\re=\delta^j_i$ we have $\le e_i,\om^j\re_\st=N^j_i$
with $N$ being a $\st$-invertible matrix since $N^j_i=\delta_i^j+O(\la)$.
We denote by $N^{-1_\st}$ the $\st$-inverse matrix of the matrix $N$.
We have $N^{-1_\st}=1+\la N_1+\la^2 N_2+\ldots$ with the
generic terms $N^{-1_\st}_n$ recursively given by 
$N^{-1_\st}_n=-\sum_{l=1}^nN^{-1_\st}_{n-l}\st N_l$, see also
\cite{G1} for another equivalent explicit expression.
Then 
\eq
\hat e_i =N^{-1_\st\,k}_{\,\,i}\st e_k~
\en
satisfies $\le \hat e_i,\om^j\re_\st=\delta^j_i\,$ as is easily seen using
$A_\st$-linearity of the pairing $\le~,~\re_\st$.
Of course we also have $\le e_i,\hat\om^j\re_\st=\delta^j_i\,$ with
$\hat\om^j =\om^k\st N^{-1_\st\,j}_{\,\,k}$. We denote by
$\{\hat\partial_\mu\}$ the basis of vectorfields that satisfy
\eq\label{vectorfieldstsdual}
\le\hat\partial_\mu , dx^\nu\re_\st=\delta_\mu^\nu~,
\en 
we have $\hat\partial_\mu=N_\mu^{-1_\st \,\nu}\st\partial_\nu$ with 
$N_\mu^\nu=\le\partial_\mu,dx^\nu\re_\st$.
\sk

Using the pairing $\le~\,,~\,\res$ we associate to any $1$-form
$\om$ the left $A_\st$-linear map $\le~\,,\om\res$. It can be shown
\cite{NG} that also the converse holds: any left $A_\st$-linear map 
$\Phi:\Xis\rightarrow \AAs$ is of the form $\le~\,,\om\res$
for some $\omega$.

\section{Covariant Derivative}\label{covderivative}
By now we have acquired enough knowledge on $\st$-noncommutative 
differential  geometry to develop the formalism of covariant
derivative,
torsion, curvature 
and Ricci tensors just by following the usual classical formalism.

We define a $\st$-covariant derivative 
$\dd^\star_u$ along the vector field $u\in \Xi$
to be a linear map $\dds_u:\Xis\rightarrow\Xis$ such that
for all $u,v,z\in\Xi_\st,~ h\in A_\st$:
\eqa
&&\dd_{u+v}^{\star}z=\dd_{u}^{\star}z+\dd_{v}^{\star}z~,\\[.35cm]
&&\dd_{h\star u}^{\star}v=h\star\dd_{u}^{\star}v~,\\[.35cm]
&&\dd_{u}^{\star}(h\star v)
\,=\,\mathcal{L}_u^{\star}(h)\star v+
\oR^\al(h)\st\dd^\st_{\oR_\al(u)}v\label{ddsDuhv}
\ena
Notice that in the last line we have used the coproduct 
formula (\ref{coproductu}),
$\D_\st (u)=u\otimes 1+ \oR_\st ^\al\otimes \ll^\st _{\oR_{\st \al}}(u)
$. Epression (\ref{ddsDuhv}) is well defined because
$\oR_\al(u)$ 
is again a vectorfield.

{\vspace{.3cm}}
\sk
\noi{\bf Local coordinates description}\\
\noi In a coordinate neighborhood $U$ with coordinates $x^\mu$ 
we have the frame $\{\hat\partial_\mu\}$ that is $\st$-dual to the
frame
$\{dx^\mu\}$ (cf. (\ref{vectorfieldstsdual})).
The (noncommutative) connection coefficients 
${\Gamma_{\mu\nu}}^\sigma$ are uniquely determined by 
\eq
\dds_{\hat\partial_\mu}\hat\partial_\nu=
{\Gamma_{\mu\nu}}^\sigma\st\hat\partial_\sigma~.
\en
They uniquely determine the connection, 
indeed for vectorfields $z$ and $u$ we have,
\eqa
\dds_z u&=&
\dds_z (u_\st^\nu\st\hat\partial_\nu)\nn\\
&=&\lls_z(u^\nu_\st)\st\hat\partial_\nu+
\oR^\al(u^\nu_\st)\st\dds_{\oR_\al(z)}\hat\partial_\nu\nn\\
&=&
\lls_z (u^\nu_\st)\st\hat\partial_\nu+
\oR^\al(u^\nu_\st)\st\oR_\al(z)^\mu\st\dds_{\hat\partial_\mu}\hat\partial_\nu\nn\\
&=&
\lls_z (u^\nu_\st)\st\hat\partial_\nu+
\oR^\al(u^\nu_\st)\st\oR_\al(z)^\mu\st
\Ga_{\mu\nu}{}^\sigma\st\hat\partial_\sigma
\ena
where $\oR_\al(z)^\mu$ are the coefficients of $\oR_\al(z)$,
$\,\oR_\al(z)=\oR_\al(z)^\mu\st\hat\partial_\mu$.
With respect to a local frame of vectorfields 
$\{e_i\}$ 
we have the 
connection coefficients
\eq\label{Connectioncoeffijk}
\dds_{e_i} e_j={\Gamma_{ij}}^k\st e_k~.
\en
\sk
\noi{\bf Covariant derivative on tensorfields}

\noi We define the covariant derivative on bivectorfields extending by
linearity the following deformed Leibniz rule, for all $u,v, z\in \Xis$,
\eq \label{Leibdds}
\dds_u(v\otimes_\st z):= \dds_{u}(v)\ots z + \oR^\al(v)\ots 
\dds_{\oR_\al(u)}z~.
\en
We now define the covariant derivative on functions to be the
$\st$-Lie derivative,
\eq
\dds_u(h)=\lls_u(h)~.
\en 
As in the commutative 
case we also define the covariant derivative on 
1-forms $\Oms$, by requiring compatibility with the contraction
operator, for all $u,v\in\Xis, \omega\in\Omega_\st$, 
\eq\label{compleredd}
\dds_u\le v,\om\res =\le\dds_u(v),\om\res + 
\le\oR^\al(v),\dds_{\oR_\al(u)} \om\res
\en
so that
$
\le v,\dds_u \om\res=
\lls_{\oR^\al(u)}\le \oR_\al(v),\om\res -\le\dds_{\oR^\al(u)}(\oR_\al(v)),\om\res\,.  
$
Finally we extend the covariant derivative to all tensorfields via 
the deformed Leibniz rule (\ref{Leibdds}) where now 
$\tau, \tau' \in \TT_\st$,
\eq\label{Leibddsg}
\dds_u(\tau\ots\tau'):= \dds_{u}(\tau)\ots 
\tau' + \oR^\al(\tau)\ots 
\dds_{\oR_\al(u)}\tau'~.
\en
\section{Torsion and Curvature}
\begin{Definition}\label{TANDR}
The torsion $\tr$ and the curvature $\rr$ associated to
a connection $\dd^\st$ are the $\mathbb{C}$-linear maps 
$\tr:\Xis\otimes_\bbC\Xis\rightarrow\Xis$, and 
$\rr^\star : \Xis\otimes_\bbC\Xis\otimes_\bbC\Xis\rightarrow\Xis$ defined by
\eqa
\tr(u,v)&:=&\dd_{u}^{\star}v-\dd_{\oR^{\alpha}(v)}^{\star}\oR_{\alpha}(u)
-[u,v]_{\star}~,\\[.2cm]
\rr(u,v,z)&:=&\dd_{u}^{\star}\dd_{v}^{\star}z-
\dd_{\oR^\al{(v)}}^{\star}\dd_{\oR_\al(u)}^{\star}z-\dds_{[u\,v]_\st} z~,
\ena
for all $u,v,z\in\Xis$.
\end{Definition}
{}From  the antisymmetry property of the bracket $[\,~]_\st$, 
see (\ref{sigmaantysymme}), and triangularity of the $\RR$-matrix it 
easily follows that 
the torsion $\tr$ and the curvature $\rr$ have the following 
$\st$-antisymmetry property
\eqa\label{Tantysymm}
\tr(u,v)&=&-\tr(\oR^\al(v),\oR_\al(u))~,\\
\rr(u,v,z)&=&-\rr(\oR^\al(v),\oR_\al(u),z)~.
\label{Rantysymm}
\ena
It can be shown \cite{NG} that  $\tr$ and $\rr$  are left $A_\st$-linear maps,
\eqa
\tr &:& \Xis\otimes_\st\Xis\rightarrow\Xis~\nn\\
\rr &:& \Xis\otimes_\st\Xis\otimes_\st\Xis\rightarrow\Xis
\ena
and therefore that they uniquely define a torsion
tensor and a curvature tensor. 
For the torsion, left $\AAs$-linearity explicitly reads
\eqa
&&\tr(f\star u,v)=f\star \tr(u,v)~,\label{trlinea1}\\
&&\tr(u,f\star v)= \tr(u\st f ,v)=\oR^\al(f)\st \tr(\oR_\al(u) ,v)~,
\label{trlinea2}
\ena 
and similarly for the curvature.
Instead of entering the technical 
Hopf algebra aspects of the proof of (\ref{trlinea1}) and (\ref{trlinea2}), we here present an easy 
intuitive argument. Recall that $f\st g=\oR^\al(g)\st\oR_\al(f)$. 
In other terms
the noncommutativity of the $\st$-product is  regulated by the
$\RR$-matrix. Expression $\oR^\al(g)\st\oR_\al(f)$ can be read as
saying that the initial ordering $f\st g$ has been
inverted. Similarly expression 
$\oR^\be\oR^\al(h)\st\oR^\be(f)\st\oR_\al(g)$ equals $f\st g\st h$
as is easily seen by 
accounting for the number of elementary transpositions
needed to permute $(f,g,h)$ into $(h,f,g)$.  In short,
$\RR^{-1}=\oR^\al\otimes\oR_\al$ is a representation of the 
permutation group on the $\st$-algebra of functions $\AAs$, 
and similarly on the algebra of vectorfields $\UU_\st$.
The formula
\eq\label{ordring1}
[f\st u\,,v]_\st=f\st[u\,,v]_\st-(\ll_{\oR^\be(\oR^\al(v))}\oR_\be(f))\st\oR_\al(u)
\en
can then be intuitively obtained recalling the analogue commutative
formula $[f u,v]=f[u,v]-(\ll_v f) u$ and keeping track of the
transpositions occurred. For example the $\RR$-matrices in the last 
addend agree with the reordering
$(f,u,v)\,\rightarrow\, (v,f,u)$.
Recalling again that the inital ordering is $(f,u,v)$ 
one similarly has 
\eq\label{ordring2}
\dds_{\oR^\al(v)}\oR_\al(f\st u)=f\st\dds_{\oR^\al(v)}\oR_\al(u)+
(\ll_{\oR^\be\oR^\al(v)}\oR_\be(f))\st\oR_\al(u)~.
\en  
The sum of (\ref{ordring1}) and of (\ref{ordring2}) gives 
the left $\AAs$-linearity property (\ref{trlinea1}) of the torsion.
Formula (\ref{trlinea2}) can be similarly obtained. It also follows
from the $\st$-antisymmetry property (\ref{Tantysymm}).
\sk\sk

\noi{\bf Local coordinates description}
\sk
\noi 
We denote by $\{e_i\}$ a local frame of vectorfields 
(subordinate to an open $U\in M$)
and by $\{\theta_j\}$ the dual frame of 1-forms:
\eq
\le e_i\,,\,\theta^j\re_\st=\delta^j_i~.
\en
The coefficients ${\tr_{ij}}^l$ and ${\rr_{ijk}}^l$ of the torsion and curvature tensors
with respect to this local frame are 
defined by 
\eqa
&&{\tr_{ij}}^l=\le\tr(e_i,e_j)\,,\,\theta^l\re_\st~,\nn\\
&&{\rr_{ijk}}^l=\le\rr(e_i,e_j,e_k)\,,\,\theta^l\re_\st~.\nn
\ena
We denote by $\La^\st$ the $\st$-transposition operator; 
it is the linear operator given by
\eq\label{transpst}
\La^\st(u\otimes_\st v)
:=\ll^\st_{\oR_\st^\al}(v)\otimes_\st\ll^\st_{\oR_{\st\al}}(u)=
{\oR^\al}(v)\otimes_\st {\oR_{\al}}(u)~.
\en
It is easily seen to be compatible with the $A_\st$-bimodule 
and the $\UU_\st$-module structure of $\Xis\ots\Xis$:
\eqa
&\La^\st(h\st u\otimes_\st v\st k)=
h\st\La^\st(u \otimes_\st v)\st k~,&\label{biliLAS}\\[.1cm]
&\ll^\st_\xi(\La^\st( u  \otimes_\st v))=
\La^\st(\ll^\st_\xi( u \otimes_\st v))~.&
\ena
Hint: use (\ref{13}),(\ref{14}), and (\ref{copopst}). 
Because of the $\AAs$-bilinearity property (\ref{biliLAS}),
we have that $\La^\st$ is completely determined by its action
on a basis of vectorfields. We define the coefficients 
$\La_{~ij}^{\st \, kl}$ of $\La^\st$
by the expression
$$\La^\st(e_i\ots e_j)=
\La_{~ij}^{\st \,kl}\st e_k\ots e_l~.
$$
Recalling the $\st$-antisymmetry property of $\tr$ and $\rr$,
(see (\ref{Tantysymm}) and (\ref{Rantysymm})), we then
immediately have the $\st$-antisymmetry properties of the 
coefficients $\tr_{ij}{}^l$ and $\rr_{nij}{}^l$,
\eq
\tr_{ij}{}^l=-\La_{~ij}^{\st \;km}\st\tr_{km}{}^l~~,~~~~
\rr_{nij}{}^l=-\La_{~ij}^{\st \;km}\st\rr_{nkm}{}^l~~.
\en
\sk
In the commutative case, if the connection is chosen to have 
vanishing torsion,  we have the first Bianchi identities
$\rr_{ijk}{}^l+\rr_{jki}{}^l+\rr_{kij}{}^l=0$, 
where the lower indices $i\,j\,k$ have been cyclically permuted.
There is a similar equation in the noncommutative case.

We first define the $\st$-operation of cyclic permutaion of three
vectors. Recalling the definition of the $\st$-transposition operator
we have that $\st$-cyclic permutation of the vectors $u\,v\,z$ 
is given by,
\eq
\mathcal{C}^\st(u\ots v\ots z)=u\ots v\ots z+\La_{12}^\st\La_{23}^\st(u\ots v\ots z)+\La_{23}^\st\La_{12}^\st(u\ots v\ots z)
\en
where $\La^\st_{12}=\La\ots id$ and $\La^\st_{23}=id \ots\La$.  
From the $\AAs$-bilinearity property of $\La^\st$ we see that also
$\CC$ is $\AAs$-bilinear
\eq
\mathcal{C}^\st(h\st u\ots v\ots z\st k)=
h\st\mathcal{C}^\st( u\ots v\ots z)\st k~.
\en
Since any tensor in $\Xis\ots\Xis\ots\Xis$ is of the form
$f^{ijk}\st e_i\ots e_j\ots e_k$, we have that the $\st$-cyclic
permutation operator is completely defined by its   
action on a basis $\{e_i\ots e_j\ots e_k\}$. 
This action is completely determined 
by the coefficients ${\mathcal{C}^\st}_{ijk}^{lmn}$ of $\CC$,
\eq
\mathcal{C}^\st(e_i\ots e_j\ots e_k)
={\mathcal{C}^\st}_{ijk}^{lmn}\st
e_l\ots e_m\ots e_n~.
\en
We can now state the  first Bianchi identity in the case of  
vanishing torsion:
\eq
\CC(\rr(u,v,z))=0
\en
where $\CC$ denotes $\st$-cyclic permutation of $u,v,$ and $z$.
In components the Bianchi identity reads
\eq
{\mathcal{C}^\st}_{ijk}^{lmn}\st\rr_{lmn}{}^p=0~.
\en
The proof of the Bianchi identity  follows the classical proof.
Since the torsion vanishes we have
$\dds_u\tr(v,z)=0$, this equation reads
\eq
\dds_u\dds_v(z)-\dds_u\dds_{\oR^\al(z)}\oR_\al(v)
-\dds_{\oR^\al([v\,,z]_\st)}\oR_\al(u)-[u\,,[v\,,z]_\st]_\st~,
\en
where we have used that $\tr(u,[v\,,z]_\st)=0$. We now add
three times this equation, 
each time $\st$-cyclically permuting 
the vectors $(u,v,z)$, so that we have the three orderings
$(u,v,z)\,$, $\,(\oR^\be\oR^\al(z),\oR_\be(u),\oR_\al(v))$ and
$(\oR^\de(v),\oR^\ga(z),\oR_\ga\oR_\de(u)).$ The three addends
$$[u\,,[v\,,z]_\st]_\st \,+ \,{\mbox{ $\st$-cyclic perm.}}$$ 
vanish because of the $\st$-Jacoby identities, 
the remaining addends give the Bianchi
identity. (This can be seen using (\ref{13}), (\ref{14}) and 
the quantum Yang-Baxter equation 
$\RR_{12}\RR_{13}\RR_{23}=\RR_{23}\RR_{13}\RR_{12}$,
that is a consequence of (\ref{12}), (\ref{13}), (\ref{14})).

\sk\sk
We end this section 
with the definition of the Ricci tensor. In the
commutative case the Ricci tensor is a contraction of the curvature
tensor, $\ric_{jk}={\rr_{ijk}}^i$.
We define the  Ricci map to be the following contraction of the
curvature:
\eq\label{defofric1}
\ric(u,v):=\le \theta^i, \rr(e_i,u,v)\res'~,
\en
where sum over $i$ is understood. The contraction
$\le~,~\res'$ is 
a contraction between forms on the {\sl left} and
vectorfields on the {\sl right}. It is defined through the by now 
familiar deformation of the commutative pairing, 
\eqa
\le \om\,,\,u\res'&:=&\le\of^\al(\om)\,,\,\of_\al (u)\re~,\\
&\,=&\le\oR^\al(u)\,,\,\oR_\al(\om)\res~.\nn
\ena
The pairing $\le~,~\res'$ has of course the 
$\AAs$-linearity properties
\eq\label{linearitypaiop}
\le h\st \om, u\st k\re_\st'=h\st\le \om, u\re_\st'\st k~~,~~~
\le \om, h\st u \re_\st'=
\le \om\st h, u\re_\st'~.
\en
Definition (\ref{defofric1}) is well given because
it is independent from the choice of the frame $\{e_i\}$ 
(and the dual frame $\{\theta^i\}$),
and because the Ricci map  so defined is an $A_\st$-linear map: 
\eqa
&&\ric(f\star u,v)=f\star \ric(u,v)~,\label{rictrlinea1}\\
&&\ric(u,f\star v)= \ric(u\st f ,v)=\oR^\al(f)\st \ric(\oR_\al(u) ,v)~.\label{rictrlinea3}
\ena 
In order to prove this statement we 
consider the coefficients $\rr^j(e_i,u,v)$ of the vector
$$\rr(e_i,u,v)=\rr^j(e_i,u,v)\st e_j~.$$ 
$\AAs$-linearity of $\rr$
implies $\AAs$-linearity of the coefficients,
$\rr^j(h\st e_i,u,v)=h\st\rr^j(e_i,u,v)$. This in turn implies 
(recall end of Section 4) 
that there exists 1-forms $\om^j_\rr(u,v)$ such that
\eq
\rr^j(e_i,u,v)= \le e_i,\om_\rr^j(u,v)\res~.
\en
From $\rr(e_i,h\st u,v)=\rr(e_i\st h, u,v)$ we immediately see that
the 1-forms $\om_\rr^j(u,v)$ are left linear in $u$, i.e.,
$\om_\rr^j(h\st u,v)=h\st\om^j_\rr(u,v)~.$
We now have
\eqa
\le \theta^i\,,\,\rr(e_i,u,v)\res'&=&
\le \theta^i\st\rr^j(e_i,u,v)\,,e_j\res'\nn\\
&=&\le \theta^i\st\le e_i,\om_\rr^j(u,v)\res\,,e_j\res'\nn\\
&=&\le \om^j_\rr(u,v)\,,e_j\res'\nn
\ena
where in the first line we used  (\ref{linearitypaiop}).
This formula implies independence from the choice of basis 
$\{e_i\}$ and left $\AAs$-linearity of $\ric$. 

The coefficients of the Ricci tensor are 
\eq
\ric_{jk}=\ric(e_j,e_k)~.
\en

\section{Metric and Einstein Equations}\label{riemgeo}
In order to define a $\st$-metric we need to define $\st$-symmetric
elements in $\Oms\ots\Oms$.
In (\ref{transpst}) we have defined the transposition operator
$\La^\st$ on vectorfields; we can similarly define it on forms,
\eq\label{transpstonforms}
\La^\st(\omega\otimes_\st \omega')
:=\ll^\st_{\oR_\st^\al}(\omega')\otimes_\st\ll^\st_{\oR_{\st\al}}(\omega)=
{\oR^\al}(\omega')\otimes_\st {\oR_{\al}}(\omega)~.
\en
We now recall that $\Oms\ots\Oms=\Om\otimes\Om$ as vectorspaces, and we notice that the transposition operator 
$\La^\st:\Oms\ots\Oms\rightarrow \Oms\ots\Oms$ 
is just the classical 
transposition operator $\La:\Om\otimes\Om\rightarrow \Om\otimes\Om$.
Indeed we have
\[
\La(\omega\otimes_\st
\omega')=\La(\of^\al(\om)\otimes\of_\al(\omega'))=
\of_\al(\om')\otimes\of^\al(\omega)=
{\oR^\al}(\omega')\otimes_\st {\oR_{\al}}(\omega)=\La^\st(\om\ots\om')~,
\] 
where in the first equality we have explicitly written the element
$\om\ots\om'$ as an element of $\Om\otimes\Om$, and then in the second equality we have applied the definition of $\La$.
This implies that (anti-)symmetric elements in $\Om\otimes\Om$ are
$\st$-(anti-)symmetric elements in $\Oms\ots\Oms$. 

Since a commutative metric is a nondegenerate symmetric tensor in
$\Om\otimes\Om\,$
we conclude that any commutative metric is also a noncommutative metric,
($\st$-nondegeneracy of the metric is insured by the fact that at
zeroth order in the deformation parameter $\la$ the metric is 
nondegenerate). Contrary to \cite{MadoreM}, \cite{MadoreFC}, 
we see that in our approach,
where all (moving) frames are on equal footing,  
there are infinitely many metrics compatible with 
a given noncommutative differential geometry, noncommutativity 
does not single out a preferred metric.

We denote by $\g$ the metric tensor.
If we write 
\eq\g=\g^a\ots\g_a\in\Oms\ots\Oms~\en 
(for example locally $\g=\theta^j\otimes_\st\theta^i\st\g_{ij}$),
then for every $v\in\Xis$ we can define the 1-form
\eq\label{defsubsid}
\le v , \g\res:=\le v,\g^a\res\st\g_a
\en
and we can then construct the left $A_\st$-linear map
$\g$, corresponding to the metric tensor $\g\in\Oms\ots\Oms$, as
\eqa\label{metricoperat}
\g\,:\,\Xis\ots\Xis &\rightarrow&\AAs\nn\\
(u,v)\,\,&\mapsto&\g(u,v)=\le u\ots v,\g\res:=
\le u\,,\le v,\g\res\res~.
\ena
The $\st$-inverse metric
$\g^{-1}\in\Xis\otimes_\st\Xis$ is then defined by the following
equations, for all $u\in\Xis, \om\in\Oms$, 
\eqa\label{thedefofmetricstinv}
\le\le u_{},_{}\g\res,\,\g^{-1}\res'=u~,\\[.1cm]
\le\le \om_{},_{}\g^{-1}\res',\,\g\res=\om~,\label{thedefofmetricstinv2}
\ena
where, as in (\ref{defsubsid}), we have defined 
\eq
\le\om , \g^{-1}\res':=\le\om,\g^{a\,-1}\res'\st\g_a^{-1}~,
\en 
and we have decomposed $\g^{-1}$ as
\eq\g^{-1}=\g^{a\,-1}\ots\g^{-1}_a\in\Xis\ots\Xis~\en
(for example locally $\g^{-1}=\g^{ij_{^{^\st}}}\st e_j\ots e_i$). 
At zeroth order in the deformation parameter $\la$, 
and using local coordinates,
we write $\g=\g_{\mu\nu}dx^\mu\otimes dx^\nu$ and the above definition
of the inverse metric gives
$\g^{-1}=\g^{\mu\nu}\partial_\mu\otimes\partial_\nu$,
where $\g^{\mu\nu}$ is the inverse matrix of $\g_{\mu\nu}$, 
$\,\g^{\mu\nu}\g_{\nu\rho}=\delta^\mu_\rho$, 
$\,\g_{\mu\nu}\g^{\nu\rho}=\delta_\mu^\rho$.
For the noncommutative analogue of the relations $\,\g^{\mu\nu}\g_{\nu\rho}=\delta^\mu_\rho$, 
$\,\g_{\mu\nu}\g^{\nu\rho}=\delta_\mu^\rho$ 
see the end of the next section.
\sk
Consider now the connection that has vanishing torsion and that
is metric compatible, $\dds_u\g=0$.  See \cite{NG}, see also \cite{G1} for  
the case $\theta$-const..  
The scalar curvature $\mathfrak{R}$ with respect to this connection 
is given by 
\eq
\mathfrak{R}:=
\ric(\g^{a\,-1},\g^{-1}_a)
\en
where $\g^{-1}=\g^{a\,-1}\ots\g^{-1}_a\in\Xis\ots\Xis$. Locally we
have $\g^{-1}=\g^{ij_{^{^\st}}}\st e_j\ots e_i$, and 
\eqa
\mathfrak{R}&=&
\ric(\g^{ij_{^{^\st}}}\st e_j\,,\, e_i)=\g^{ij_{^{^\st}}}\st \ric(e_j,
e_i)\nn\\&=&\g^{ij_{^{^\st}}}\st \ric_{ji}~.
\ena
We finally arrive at the noncommutative Einstein equation (in vacuum),
\eq\label{Einstein}
\ric-{1\over 2}{_{_{}}}\g\st\mathfrak{R}=0~,
\en
where the dynamical field is the metric $\g$.
This equation is an equality between the left $\AAs$-linear
maps $\ric$ and $\g\st\mathfrak{R}$,  where
$$(\g\st\mathfrak{R})(u,v):=\le u\ots v , \g\st\mathfrak{R}\res=
\le u\ots v , \g\res\st\mathfrak{R}=
\g(u,v)\st\mathfrak{R}~.$$
Because of left $A_\st$-linearity the curvature scalar 
must appear on the right of the metric and not on the left
in (\ref{Einstein}). 
Applying (\ref{Einstein}) to the vectors $e_i$ and $e_j$ we obtain the
components equation
\eq
\ric_{ij}-{1\over 2}{_{_{}}}\g_{ij}\st\mathfrak{R}=0~,
\en
where $\g_{ij}=\g(e_i,e_j)=\le e_i\ots e_j,\g\res\,$ are the same
coefficients appearing in the expression $\g=\theta^j\otimes_\st\theta^i\st\g_{ij}$.

\section{Conjugation}
In this section we introduce the notion of complex conjugation on the
algebra $\AAs$, and we see that we can impose reality conditions on the
$\st$-spaces of functions, vectorfields and tensorfields.

\sk
We first briefly recall the commutative $\ste$-structure. Given a smooth real 
manifold $M$, the usual $\ste$-structure on the complex valued functions 
$A=Fun(M)$ is a map $\ste :A\rightarrow A\,,$
 where for all $h\in A$ and  $m\in M$,
\eq
h^\ste(m)=\overline{h(m)}~,
\en
here the bar $\,\overline{^{^{~~}}}\,$ denotes complex conjugation.
This $\ste$-structure induces a $\ste$-structure on the Lie algebra of 
vectorfields by defining $\ste : \Xi\rightarrow \Xi$, where for all $u\in\Xi$
and $h\in A$, 
\eq\label{qwer}
u^\ste(h):=(S(u)(h^\ste))^\ste=-(u(h^\ste))^\ste~.
\en
It is easy to check that the $\ste$-operation so defined is antimultiplicative with respect to the Lie bracket of $\Xi$, $[u,v]^\ste=[v^\ste,u^\ste]$.
In particular, locally, we can consider the real coordinate functions $x^\mu$,
then the partial derivatives $\partial_\mu$ are pure imaginary, $\partial^\ste_\mu=-\partial_\mu$; we also have $u^\ste=(u^\mu\,\partial_\mu)^\ste=
-\overline{u^\mu}\,\partial_\mu$.

The $\ste$-structure 
on $\Xi$ is extended to the universal enveloping algebra $\UU$ by antilinearity and antimultiplicativity, so that for all $\xi,\zeta\in \UU$, $(\xi\zeta)^\ste=\zeta^\ste\xi^\ste$. 
Applying a vectorfield $v$ to definition (\ref{qwer})
we obtain $(v^\ste u^\ste)(h)=(S(uv)(h^\ste))^\ste$, and iterating we obtain that for a generic element of $\UU$,
\eq
\xi^\ste(h)=(S(\xi)(h^\ste))^\ste~.
\en
Similarly from $u^{\,\ste}(v)=[u^\ste,v]=[S(u),v^\ste]^\ste=(S(u)(v^\ste))^\ste$ we have,
\eq\label{relvform}
\xi^\ste(\zeta)=(S(\xi)(\zeta^\ste))^\ste~.
\en
Finally, from the local formula $\le \partial_\mu, dx^\nu\re^\ste=-\le \partial^\ste_\mu, (dx^\nu)^\ste\re^\ste$ we have the general formula of compatibility 
between the $\ste$-structure and the pairing
\eq
\le u,\om\re^\ste=-\le u^\ste , \om^\ste\re~.
\en

\sk
We now study the $\ste$-operation in the noncommutative context.
We define the $\ste$-structure on $\AAs$ to be the same as that on $A$. The requirement 
\eq
(h\st g)^\ste=g^\ste\st h^\ste
\en
is then satisfied if the twist $\FF$ satisfies the relation 
$(S\otimes S)\FF_{21}=\FF^{\ste\otimes\ste}$, i.e.,
\eq
(S\otimes S)\FF^{-1}_{21}={\FF^{-1}}^{\ste\otimes\ste}~.
\en

We similarly define the $\ste$-structure on 
$\UU$ to be the same as the undeformed one. Using (\ref{relvform}) 
it is not difficult to show that the $\ste$-operation is compatible 
with the $\st$-product of $\UU_\st$ and with the $\st$-Lie bracket of $\Xis$,
\eq
(\xi\st\zeta)^\ste=\zeta^\ste\st\xi^\ste~~~,~~~~~~
[u,v]_\st{}^\ste=[v^\ste,u^\ste]_\st~.
\en
It can be shown \cite{NG} that the $\ste$-operation  
is compatible with the triangular Hopf algebra structure of $\UU_\st$
(a key point being that on $\UU^\FF$ the $\ste$-operation reads 
$\xi^{\ste_\FF}:=\chi\xi^\ste\chi^{-1}$).
On tensors too the $\ste$-structure is by definition 
the undeformed  one, and we have,
for all $\tau, \tau'\in \TT_\st$, 
\eq\label{steontau}
(\tau\ots\tau')^\ste=\oR^{\al}(\tau^\ste)\ots\oR_\al(\tau'^\ste)~.
\en
Finally the two pairings $\le~,~\res$ and $\le~,~\res'$ are related by the 
$\ste$-operation, for all $u\in\Xis$ and $\om\in\Oms$, we have
\eq
\le u , \om\res^{\;\ste}=-\le\om^\ste, u^\ste\res'~.
\en
In particular, if 
locally we consider a basis $\{e_i\}$ and the dual basis $\{\theta^i\}$,
\[
\le e_i,\theta^j\res=\delta^j_i~,
\]
then the $\ste$-conjugate basis $\{e_i^\ste\}$ and $\{\theta^{j\,\ste}\}$ 
are (up to a sign) dual with respect to the $\le~,~\res'$ pairing,
\eq
\le \theta^{j\,\ste} , e_i^{\,\ste}\res'=-\delta^j_i~.
\en
\sk
We can now study for example the reality property 
\eq
\g^\ste=\g
\en 
of the
metric tensor $\g\in \Oms\ots\Oms$.
The metric tensor has a convenient expansion in terms of the
$\theta^i$ and the $\theta^{\bar\jmath\,\ste}$ 1-forms (here $\bar\jmath$ is just an index like $i$ or $j$). We set 
\eq\label{exp153}
\g
=\theta^i\ots\g_{i\bar\jmath}\st\theta^{\bar\jmath\,\ste}~.
\en
In this basis reality of the metric, and therefore of
the noncommutative  Einstein equations, has a very simple explicit expression. 
Also the explicit expression for the inverse metric is particularly
simple in this basis.

We first study the consequences of the reality condition $\g=\g^{\ste}$
on the metric coefficients $\g_{i\bar\jmath}$. 
From (\ref{steontau}) we have,
\eq\label{grtrtg}
\g^\ste=\oR^\al(\theta^{i\,\ste})\ots\oR_\al
(\theta^{\bar\jmath}\st\g_{i\bar\jmath}^\ste)=
\oR^\al(\theta^{\bar\jmath\,\ste})\ots\oR_\al
(\theta^{i}\st\g_{\bar\jmath i}^\ste)
\en
where in the last equality we have just renamed the indices.
In order to compare this expression of $\g^\ste$ with the expression 
(\ref{exp153}) of $\g$,
we  use the $\st$-symmetry property of the metric,
$\g=\La^\st\g$, to rewrite the metric as
$$\g=\theta^i\st\g_{i\bar\jmath}\ots\theta^{\bar\jmath\,\ste}=
\oR^\al(\theta^{\bar\jmath\,\ste})\ots\oR_\al(\theta^i\st\g_{i\bar\jmath})~.$$
Comparison with (\ref{grtrtg}) gives, 
$
\oR^\al(\theta^{\bar\jmath\,\ste})\ots\oR_\al
(\theta^{i}\st\g_{\bar\jmath i}^\ste)
=\oR^\al(\theta^{\bar\jmath\,\ste})\ots\oR_\al(\theta^i\st\g_{i\bar\jmath})\,
$
iff $\g=\g^\ste$. After applying the transposition $\La^\st$ to this
equation we obtain that reality of $\g$ reads
$$
\theta^{i}\st\g_{\bar\jmath i}^\ste\ots
\theta^{\bar\jmath\,\ste}
=
\theta^i\st\g_{i\bar\jmath}\ots\theta^{\bar\jmath\,\ste}~,
$$
i.e.,
\eq
\g_{\bar\jmath i}^\ste=\g_{i\bar\jmath}~.
\en

Concerning the inverse metric $\g^{-1}$, we have that it is given by
the expression
\eq\label{gmen1}
\g^{-1}=-e_{\bar\imath}^\ste\ots\g^{\bar\imath j_{^\st}}\st e_j
\en 
where $\g^{\bar\imath j _{^\st}}$ is the $\st$-inverse matrix of
$\g_{i\bar\jmath}$, 
$$\g^{\bar\imath j _{^\st}}\st\g_{j\bar\ell}=\delta_{\bar\ell}^{\bar\imath}\,~~,~~~~
\g_{i\bar\jmath}\st \g^{\bar\jmath
  \ell _{^\st}}=\delta_{i}^{\ell}~.$$
Indeed it is not difficult to see that (\ref{gmen1}) satisfies
(\ref{thedefofmetricstinv}) and (\ref{thedefofmetricstinv2}). 
\sk
\sk
\noi{\bf{\large Acknowledgements}}
\sk
\noi We are very thankful to Christian Blohmann and Peter Schupp for 
many useful discussions and insights. 
Part of this work is based on our common work \cite{G1}.
We would also like to thank Gaetano Fiore, 
John Madore and Stefan Waldmann for their comments. Hospitality and
support from
Arnold Sommerfeld Center for Theoretical Physics and 
Max-Planck-Institut f\"ur Physik M\"unchen, and from
 INFN Torino and  Universit\'a di Torino (MIUR contract
PRIN 2003023852) is acknowledged.

\sk

\appendix
\section{Appendix}

\subsection{Proof that $\UU^\FF$ is a  Hopf Algebra}\label{Hopf F}
We start from
\begin{equation}
(\varepsilon \otimes id)\Delta^{\cal F}(u) = u = (id \otimes \varepsilon)\Delta^{\cal F}(u)  \label{2.4.5}
\end{equation}
and calculate first the left hand side
\begin{eqnarray}
(\varepsilon \otimes id)\Delta^{\cal F}(u) &=& (\varepsilon \otimes id)
(\f^\alpha u_1 \bar{\f}^\beta \otimes \f_\alpha u_2 \bar{\f}_\beta) \nonumber\\
&=& \varepsilon(\f^\alpha u_1 \bar{\f}^\beta) \f_\alpha u_2 \bar{\f}_\beta 
= \varepsilon(\f^\alpha)\varepsilon(u_1) \varepsilon(\bar{\f}^\beta) \f_\alpha u_2 \bar{\f}_\beta ~.\label{2.4.5'}
\end{eqnarray}
In the last line we have used that $\varepsilon : U\Xi \to \mathbb{C}$ is an algebra homomorphism.
Applying $(\varepsilon\otimes id)$ on the identity
\begin{equation}
{\cal F}{\cal F}^{-1} = 1 \otimes 1 \label{2.4.7}
\end{equation}
and using (\ref{2.23}) gives
\begin{eqnarray}
1=(\varepsilon\otimes id) {\cal F}{\cal F}^{-1} &=&
(\varepsilon\otimes id) (\f^\alpha\bar{\f}^\beta\otimes \f_\alpha\bar{\f}_\beta) \nonumber\\
&=& \varepsilon(\f^\alpha)\varepsilon(\bar{\f}^\beta)\f_\alpha\bar{\f}_\beta 
= \varepsilon(\bar{\f}^\beta)\bar{\f}_\beta ~.\label{2.4.8}
\end{eqnarray}
Inserting this into (\ref{2.4.5'}) 
we finally obtain
\begin{equation}
(\varepsilon \otimes id)\Delta^{\cal F}(u) = \varepsilon(u_1)u_2 = u~ .\label{2.4.5''}
\end{equation}
In order to calculate the right hand side of (\ref{2.4.5}) one proceeds in the analogous way.

Next we prove
\begin{equation}
\mu (S^{\cal F}\otimes id )\Delta^{\cal F}(u) = \varepsilon(u)1 
= \mu( id \otimes S^{\cal F})\Delta^{\cal F}(u)\, .\label{2.4.9}
\end{equation}
To show this we first have to prove that $\chi^{-1}=S(\of^\alpha)\of_\alpha\,$:
\begin{eqnarray}
\chi \chi^{-1}& =& \f^\beta S(\f_\beta)S(\of^\alpha)\of_\alpha \nn\\  
&=& \of^\gamma \varepsilon(\of_\gamma)\f^\beta S(\of^\alpha\f_\beta)\of_\alpha \nn \\ 
&=& \of^\gamma \f^\beta S(\of_{\gamma_{1}}\of^\alpha \f_\beta)\of_{\gamma_{2}}\of_\alpha \nn \\
&=& \of^\gamma_{1} S(\of^\gamma_{2}))\of_{\gamma} \nn \\
&=& \varepsilon(\of^\gamma) \of_{\gamma} =1\,. \nn 
\end{eqnarray}
In the first line we used the definitions given in (\ref{2.4.2}). Next we inserted $1=\of^\gamma \varepsilon(\of_\gamma)$ which we showed in (\ref{2.4.8}).  
The antipode property $S(\xi_{1})\xi_{2}=\varepsilon(\xi)$ together with the fact that the antipode is an antialgebra homomorphism lead to the next line. Then we used $ \of^\gamma \f^\beta \otimes  \of_{\gamma_{1}}\of^\alpha \f_\beta \otimes \of_{\gamma_{2}}\of_\alpha=\of^\gamma_{1} \otimes \of^\gamma_{2} \otimes \of_{\gamma}$ which follows from the cocycle condition (\ref{ass}) by multiplying both sides of the equality with $\f^\beta \otimes \f_\beta \otimes 1$. The next step uses  the antipode property $\xi_{1}S(\xi_{2})=\varepsilon(\xi)$. Finally we used $\varepsilon(\of^\gamma) \of_\gamma=1$. Similarly one shows that $\chi^{-1}\chi=1$.

We are now able to prove (\ref{2.4.9}). Starting with the left hand side we get

\begin{eqnarray}
\mu (S^{\cal F}\otimes id )\Delta^{\cal F}(u) &=& \mu \big( S^{\cal F}(u_{1_{\cal F}})\otimes u_{2_{\cal F}} \big) \nonumber\\
&=& \f^\alpha S(\f_\alpha)S(\f^\gamma u_1 \bar{\f}^\delta)S(\bar{\f}^\beta)\bar{\f}_\beta 
\f_\gamma u_2 \bar{\f}_\delta \nonumber\\
&=& \f^\alpha S(\f_\alpha)S(\bar{\f}^\beta \f^\gamma u_1 \bar{\f}^\delta)
\bar{\f}_\beta  \f_\gamma u_2 \bar{\f}_\delta \nonumber\\
&=& \f^\alpha S(\f_\alpha)S(u_1 \bar{\f}^\delta)u_2 \bar{\f}_\delta \label{2.4.9'}\nn\\
&=&\f^\al S(f_\al)S(\of^\delta) S(u_1)u_2\of_\delta~.
\end{eqnarray}
Here we used that $S$ is an antialgebra homomorphism and that $\FF\FF^{-1}=\of^\be\of^\ga\otimes\of_\be\of_\ga=1\otimes 1\,.$

Knowing that $\Delta$ is the coproduct in the $U\Xi$ Hopf algebra we find
\begin{equation}
\mu (S\otimes id )\Delta(u) = S(u_1)u_2 = \varepsilon(u)\,. \label{2.4.11}
\end{equation}
Inserting this relation into (\ref{2.4.9'}) gives
\begin{eqnarray}
\mu (S^{\cal F}\otimes id )\Delta^{\cal F}(u) 
&=& \f^\alpha S(\f_\alpha)S(\bar{\f}^\delta)\varepsilon(u)\bar{\f}_\delta\nonumber\\
&=& \chi \chi^{-1}\varepsilon(u) = \varepsilon(u)\,. \label{2.4.9''}
\end{eqnarray}
The right hand side of (\ref{2.4.9}) one proves analogously.

\subsection{$\st$-Jacoby identity}\label{AJacoby}
In order to prove the $\st $-Jacobi identity, 
$
[u \,[v\,z]_\st ]_\st =[[u\,v]_\st \,z]_\st  + [\oR^\al(v)\,\, [\oR_\al(u)\,\,z]_\st ]_\st 
$ we use the following
\sk
\noi {\bf{Lemma}}
\eq\label{lemma}
\of^\al\oR^\ga\otimes\of_{\al_1}\of^\be\oR_\ga\otimes
\of_{\al_2}\of_\be=\of^{\al}_{~_1}\of_\de\otimes\of^{\al}_{~_2}\of^\de\otimes
\of_{\al}
\en
\begin{proof}
\eqa
\of^\al\oR^\ga\otimes\of_{\al_1}\of^\be\oR_\ga\otimes
\of_{\al_2}\of_\be &=&
\of^\al\f^\ga\of_\de\otimes\of_{\al_1}\of^\be\f_\ga\of^\de\otimes
\of_{\al_2}\of_\be \nonumber \\
&=&
\of^{\al}_{~_1}\of^\be\f^\ga\of_\de\otimes\of^{\al}_{~_2}\of_\be\f_\ga\of^\de\otimes
\of_{\al}\nonumber \\
&=&\of^{\al}_{~_1}\of_\de\otimes\of^{\al}_{~_2}\of^\de\otimes
\of_{\al} 
\nonumber
\ena
where in the third line we applied  property (\ref{ass}), 
while in the last line
we used that $\of^\be\f^\ga\otimes \of_\be\f_\ga=\FF^{-1}\FF=1\otimes 1$.
\end{proof}
Now we observe that $\forall \xi \in\UU$,
\eq 
\ll_\xi([v\,z])= 
\ll_\xi (vz)- \ll_\xi( zv) =\ll_{\xi_1}( v) \ll_{\xi_2} (z)- \ll_{\xi_1} (z) \ll_{\xi_2}(v) 
=[\ll_{\xi_1}(v)\,\ll_{\xi_2}(z)]
\en 
where we used  $\ll_{\xi_1} (z) \ll_{\xi_2}(v) = \ll_{\xi_2}( z)
\ll_{\xi_1}(v) $
which holds because the classical coproduct $\Delta$ (see (\ref{cosclass})) is
cocommutative. Finally we have the $\st$-Jacoby identity
\eqa
[u \,\,[v\,z]_\st ]_\st &=&
[\of^\al(u)\,\,[\of_{\al_1}\of^\be(v)\,\,\of_{\al_2}\of_\be(z)]]\nonumber\\
&=&
[\of^{\al}_{~_1}\of^\be(u)\,\,[\of^{\al}_{~_2}\of_\be(v)\,\,\of_{\al}(z)]]\nonumber\\
&=&
[[\of^{\al}_1\of^\be(u)\,\,\of^{\al}_2\of_\be(v)]\,\,\of_{\al}(z)]+
[\of^{\al}_{~_2}\of_\be(v)\,\,[\of^{\al}_{~_1}\of^\be(u)\,\,\of_{\al}(z)]]
\nonumber\\
&=&
[[u\,v]_\st \,\,z]_\st  + [\oR^\al(v)\,\, [\oR_\al(u)\,\,z]_\st ]_\st 
\ena
where in the second line we used property (\ref{ass}), while in the last line
we used the above lemma and the fact that $\UU$ is cocommutative.

\subsection{Associativity of the $\st$-product on superspace}\label{susyass}
First we calculate
\begin{eqnarray*}
(g\star h)\star k & = & \mu\circ\mathcal{F}^{-1}(g\star h\otimes k)\\
 & = & \mu\circ\mathcal{F}^{-1}((\mu\circ\mathcal{F}^{-1}(g\otimes h))\otimes k)\\
 & = & \mu\circ\mathcal{F}^{-1}\circ((\mu\circ\mathcal{F}^{-1})\otimes id)(g\otimes h\otimes k)\\
 & = & \mu\circ\mathcal{F}^{-1}\circ(\mu\otimes id)\circ(\mathcal{F}^{-1}\otimes id)(g\otimes h\otimes k)\\
 & = & \mu\circ(\mu\otimes id)\circ(\Delta\otimes id)\mathcal{F}^{-1}\circ(\mathcal{F}^{-1}\otimes id)(g\otimes h\otimes k)\\
 & = &
 \mu\circ(\mu\otimes id)\circ\left(((\Delta\otimes id)\mathcal{F}^{-1})\mathcal{F}_{12}^{-1}\right)(g\otimes h\otimes k)~,
\end{eqnarray*}
 where 
in the last line we used that $\ll_\xi\circ\ll_\zeta=\ll_{\xi\zeta}$
(i.e. $\xi\circ\zeta(h)=\xi\zeta(h)\,$), and 
in the next to last line we used that 
\begin{eqnarray*}
\mathcal{F}^{-1}\circ(\mu\otimes id)(g'\otimes h'\otimes k') & = & (-1)^{{|\of_{\alpha}|}|g'h'|}\of^{\alpha}(g'h')\otimes\of_{\alpha}(k')\\
 & = & (-1)^{{|\of_{\alpha}|}|g'h'|+|\of_{2}^{\alpha}||g'|}\of_{1}^{\alpha}(g')\of_{2}^{\alpha}(h')\otimes\of_{\alpha}(k')\\
 & = & (\mu\otimes id)\circ(\Delta\otimes id)\mathcal{F}^{-1}(g'\otimes h'\otimes k')~.\end{eqnarray*}
Then we similarly obtain \begin{eqnarray*}
g\star(h\star k) & = & \mu\circ\mathcal{F}^{-1}(g\otimes(h\star k))\\
 & = & \mu\circ\mathcal{F}^{-1}(g\otimes(\mu\circ\mathcal{F}^{-1}(h\otimes k)))\\
 & = & \mu\circ\mathcal{F}^{-1}\circ(id\otimes\mu)\circ(id\otimes\mathcal{F}^{-1})(g\otimes h\otimes k)\\
 & = & \mu\circ(id\otimes\mu)\circ\left(
   ((id\otimes\Delta)\mathcal{F}^{-1})\mathcal{F}_{23}^{-1}\right)(g\otimes h\otimes k)~.\end{eqnarray*}
 Using (\ref{ifpppop}) we finally conclude that 
$(g\star h)\star k=g\star(h\star k)$, and associativity is proven.

\end{document}